\documentclass[twocolumn]{aastex63}
\received{April 16, 2019}
\accepted{January 15, 2021}
\submitjournal{ApJ}

\shorttitle{3-D simulations for Proxima b detectability}
\shortauthors{Galuzzo et al.}

\begin{document}

\title{3-D climate simulations for the detectability of Proxima Centauri b}

\correspondingauthor{Luca Giovannelli}
\email{luca.giovannelli@roma2.infn.it}

\author{Daniele Galuzzo}
\affil{Department of Physics, University of Rome Tor Vergata, Via della Ricerca Scientifica, 1, I-00133, Rome, Italy}

\author{Chiara Cagnazzo}
\affiliation{Institute of Marine Sciences (ISMAR), 
National Research Council (CNR), Via del Fosso del Cavaliere, 100, I-00133, Rome, Italy}

\author{Francesco Berrilli}
\affil{Department of Physics, University of Rome Tor Vergata, Via della Ricerca Scientifica, 1, I-00133, Rome, Italy}

\author{Federico Fierli}
\affiliation{Institute of Atmospheric and Climate Sciences (ISAC), 
National Research Council (CNR), Via del Fosso del Cavaliere, 100, I-00133, Rome, Italy}

\author{Luca Giovannelli}
\affil{Department of Physics, University of Rome Tor Vergata, Via della Ricerca Scientifica, 1, I-00133, Rome, Italy}



\begin{abstract}

The discovery of a planet orbiting around Proxima Centauri, the closest star to the Sun, opens new avenues for the remote observations of the atmosphere and surface of an exoplanet, Proxima b. To date, three-dimensional (3D) General Circulation Models (GCMs) are the best available tools to investigate the properties of the exo-atmospheres, waiting for the next generation of space and ground-based telescopes.
In this work, we use the PlanetSimulator (PlaSim), an intermediate complexity 3D GCM, a flexible and fast model, suited to handle all the orbital and physical parameters of a planet and to study the dynamics of its atmosphere. Assuming an Earth-like atmosphere and a 1:1 spin/orbit configuration (tidal locking),
our simulations of Proxima b are consistent with a day-side open ocean planet with a superrotating atmosphere.
Moreover, because of the limited representation of the radiative transfer in PlaSim, we compute the spectrum of the exoplanet with an offline Radiative Transfer Code with a spectral resolution of 1 nm. This spectrum is used to derive the thermal phase curves for different orbital inclination angles. In combination with instrumental detection sensitivities, the different thermal phase curves are used to evaluate observation conditions at ground level (e.g., ELT) or in space (e.g., JWST).
We estimated the exposure time to detect Proxima b (assuming an Earth-like atmosphere) thermal phase curve in the FIR with JWST with signal-to-noise ratio $\simeq$1. Under the hypothesis of total noise dominated by shot noise, neglecting other possible extra contribution producing a noise floor, the exposure time is equal to 5 hours for each orbital epoch.

\end{abstract}


\section{Introduction} \label{sec:intro}

A planet in the habitable zone of Proxima Centauri was detected in \cite{Anglada2016:paper} and confirmed recently in \cite{Damasso2020} and \cite{Suarez2020}.
The planet, known as Proxima b, has not been observed as a transiting planet \citep{no-transit1,no-transit2,no-transit3,no-transit4,no-transit5,no-transit6} 
and the estimated geometric probability of a transit is about 1.5\%, (\citealt{Anglada2016:paper}).
As a consequence, constraints can exclusively be set on the orbit semi-major axis and on the planet's revolution period, leaving mass, radius, and density of the planet to be assumed.\\
Such an indeterminacy on the physical parameters as well as on the chemical composition of a possible atmosphere or ocean/land distribution, makes every hypothesis on the climate of Proxima b extremely speculative.
Nevertheless, thanks to the proximity of Proxima b, remote observations of its atmosphere and its surface will be possible in the next decade (see e.g. \citealt{Kuhn2018:paper}). 

Nevertheless, some authors started assessing the Proxima b observation feasibility with current or forthcoming telescopes, investigating the possible detection limits.
\cite{Turbet2016:paper} used a 3D General Circulation Model (GCM), derived from LMDZ (\citealt{Hourdin2006:paper}) and assuming an Earth-like atmosphere to compute synthetic emission spectra of the planet, proposed for the first time to observe Proxima b with the James Webb Space Telescope (JWST), either with direct imaging or with thermal phase curves.
\cite{Kreidberg2016:paper}, using an analytic \textit{toy climate model} to presume the planet thermal phase curve, estimated in more details the JWST noise, and how the infrared thermal phase curve could be used to reveal the presence of an atmosphere on Proxima b.
\cite{Lovis2017:paper} proposed a theoretical setup in which Proxima b could be observed with a 8-m class telescope through a high resolution high contrast technique.
Results similar to \cite{Turbet2016:paper} were obtained by \cite{Boutle2017:paper} using the Met Office Unified Model (\citealt{Walters2017:paper}), which has a higher spectral resolution than LMDZ.
In this work, we propose a new set-up of simulations for the Proxima b atmosphere performed with an intermediate complexity, flexible and fast 3D GCM model integrated with an offline 1D Radiative Transfer Code (RTC).
Specifically, the 3D GCM is a modified version of the Planet Simulator (PlaSim, \citealt{Fraedrich2005a:paper}) developed at University of Hamburg and based on the Reading multi-level spectral Simple Global Circulation Model (SGCM) described by \cite{Hoskins1975:paper}. This model, developed to maximize the compatibility with the comprehensive European Centre Hamburg (ECHAM) GCM, has been applied and tested in different research fields, including: climate change and variability studies (\citealt{Lunkeit1998:paper}, \citealt{Bordi2007:paper} and \citealt{2015AnGeo..33..267B}), atmospheric dynamics and thermodynamics theoretical studies (\citealt{PerezMunuzuri2005:paper}, \citealt{Seiffert2007:paper}, \citealt{Kunz2009:paper}, \citealt{Schmittner2011:paper}, \citealt{Bordi2012a:paper} and \citealt{Fraedrich2012:paper}), sensitivities studies for Earth climate (\citealt{Fraedrich2005b:paper}, \citealt{Romanova2006:paper}, \citealt{Lucarini2010a:paper}, \citealt{Bordi2012b:paper}, \citealt{Bathiany2012:paper} and \citealt{Knietzsch2015:paper}), simulations of Solar System planetary atmosphere (\citealt{Grieger2004:paper}, \citealt{Segschneider2005:paper} and \citealt{Stenzel2007:paper}) and simulations of planetary atmosphere in general (\citealt{Lucarini2010b:paper}, \citealt{Lucarini2013:paper}, \citealt{Pascale2013:paper}, \citealt{Boschi2013:paper}, \citealt{Linsenmeier2015:paper}, \citealt{Gomezleal2018:paper} and \citealt{Gomezleal2019:paper}). 
PlaSim is a fast and flexible model, adaptable to extensively alter planetary parameters, but, at the same time, it takes into account as many processes occurring on Earth's atmosphere as possible. 
However, PlaSim only provides the integrated planetary atmosphere radiation within three bands, two for the shortwave radiation and one for the longwave radiation.
Thus, to explore the planet emission spectra and to derive thermal phase curves, we combine the RTC \textit{uvspec} to PlaSim. \textit{uvspec} is included in the libRadTran library (\citealt{Emde2016:paper}) and uses the \textit{DIScrete Ordinate Radiative Transfer} (DISORT) solver, reviewed by \cite{Stamnes1988:paper} and \cite{Tsay2000:paper}. In this work, \textit{uvspec} is used in an offline configuration, i.e. it is not directly implemented within the PlaSim code, but it runs on PlaSim outputs: the atmospheric state (e.g. trace gas profiles, temperature and pressure profiles, surface properties and eventually cloud liquid water content, cloud droplet size) obtained by the PlaSim 3D simulation is given as input to \textit{uvspec} for each atmospheric vertical column. Our approach, in particular the offline radiative transfer calculations, provides robust results which are in line with those obtained with more complex models. Furthermore, libRadTran model allows us to have an accurate representation of the atmospheric emission spectra, useful to estimate the signal-to-noise ratio and evaluate the detection limits of broadband planetary emission for different orbital configurations. \\
This article is structured as follows. In Section \ref{sec:system_settings} we present the known properties and parameters of the Proxima system. In Section \ref{sec:models} we describe the interactions between PlaSim and \textit{uvspec} and we discuss the assumptions and parameterizations used in our simulation. 
In Section \ref{sec:Proxima_b_climate} the dynamical properties of an Earth-like atmosphere for Proxima b are reported, assuming its minimum mass as the planet mass.
In Section \ref{sec:detectability}, we evaluate the importance of planetary atmosphere and climate when assessing the observational limits through the use of thermal phase curves technique.
Finally, in Section \ref{sec:conclusions} we report our conclusions.

\section{Inputs for the simulation of Proxima Centauri system} \label{sec:system_settings}

\subsection{Proxima Centauri stellar spectral irradiance} \label{subsec:star_parameters}

\begin{figure*}[htp]
  \centering
  \includegraphics[width=350pt]{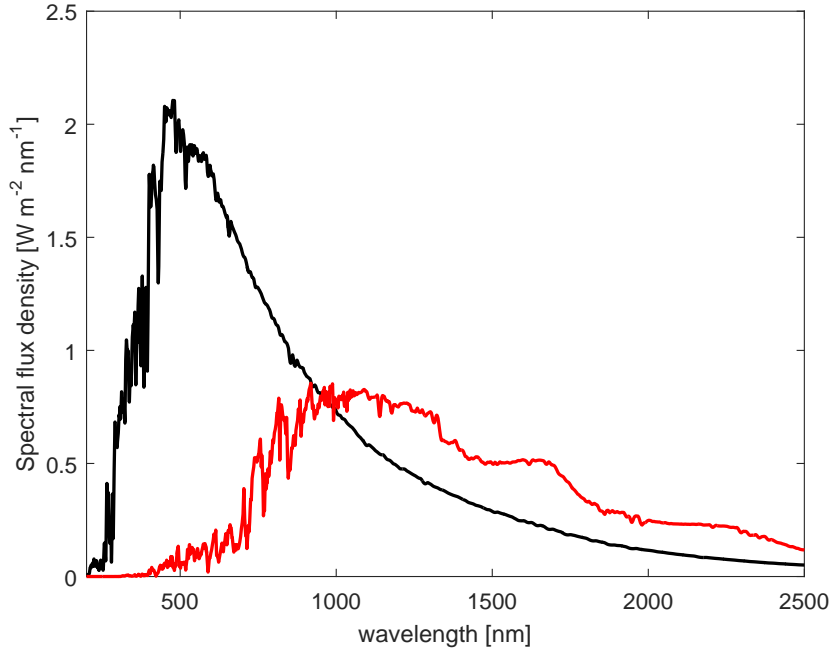}
  \caption{Sun (black curve, \url{http://kurucz.harvard.edu/}) and Proxima Centauri (red curve, \citealt{Meadows2018}) spectral irradiances as measured and evaluated at the top of the atmosphere of Earth and Proxima b, respectively.}
  \label{Fig:Spectra_comparison}
\end{figure*}

Proxima Centauri is the Solar System nearest star with a distance of $4.244 \pm 0.001$ light-years \citep{gaia_dr2_2018}. It is a red dwarf of spectral type M5.5e (\citealt{Bessel1991:paper}), with estimated radius and mass $R_{\star} = 0.141 \pm 0.021 \; R_{\odot}$ and $M_{\star} = 0.120 \pm 0.015 \; M_{\odot}$ (\citealt{Anglada2016:paper}), respectively. The effective temperature estimated by \cite{Ribas2017:paper} is $T^{eff}_{\star} = 2980 \pm 80 \; \mathrm{K}$. Proxima Centauri peak of emission is in the Near-Infrared region and the spectral energy distribution is largely different from a G-type, Sun-like star. A comparison between Sun and Proxima Centauri spectral energy distributions is shown in Figure \ref{Fig:Spectra_comparison}, where the star fluxes are evaluated at the top-of-the-atmosphere (TOA) of Earth and the planet Proxima b, respectively.

Proxima Centauri spectrum is obtained following \citealt{Meadows2018} and using the version 2.2 of the MUSCLES Treasury Survey data products, discussed in \cite{France2016:paper}, \cite{Youngblood2016:paper} and \cite{Loyd2016:paper}. The total irradiance of Proxima Centauri used in this work, which was obtained by integrating the red curve in Figure \ref{Fig:Spectra_comparison}, is equal to $885 \pm 44 \; \mathrm{W \, m^{-2}}$.
This value is similar to that of \cite{Boutle2017:paper} and \cite{DelGenio2019:paper}, equal to $881.7 \; \mathrm{W \, m^{-2}}$, while it turns out to be about 8\% less than the irradiance proposed by \cite{Turbet2016:paper}, as already discussed by \cite{Boutle2017:paper}. 
Recently, observations of Proxima b compatible with super-flare occurrence, probably produced by magneto-convective processes similar to those observed on a small scale in our sun, have been reported \citep{Howard2018:paper}. We know that in the solar case the magnetic activity of the star causes variations in irradiance. Small variations ($ \simeq 0.1 \% $) are reported in the case of total solar irradiance (i.e. bolometric irradiance), but these can lead to considerable variations ($\simeq 10-100 \% $) in particular spectral ranges, for example in the EUV or UV \citep[e.g.][]{Criscuoli2019:paper}.\\
We therefore expect that Proxima b also has a variable luminosity, both bolometric and spectral.
However, \cite{no-transit6} reported  Spitzer Space Telescope observations of Proxima Centauri that showed reduced stellar activity at near-to-mid infrared spectral range with respect to the optical one.
In this paper, for simplicity, we will assume a constant star luminosity, i.e., a constant stellar spectral flux.

The presence of planetary atmospheres results from complex mechanisms involving stellar wind, flares, Coronal Mass Ejections, and the presence of a planetary magnetic field or other shielding effects. \cite{Zuluaga2013:paper} suggested that Proxima b may have a planetary magnetic field if the planet has an Earth-like interior. The expected high dynamic pressure of Proxima Centauri wind pose hard constraints on the possibility for Proxima b to host an atmosphere on Gyr timescale. Nevertheless, the presence of a magnetosphere can help to conserve an exoplanet's atmosphere by reducing the erosion effects from stellar winds \citep[e.g.][]{Seeetal2014:paper,Dong2017}. Furthermore, the atmospheric escape rate crucially depends on many parameters (e.g. stellar wind properties, composition and thickness of the atmosphere). Here, we assume an Earth-like atmosphere with oxic composition. 

\subsection{Proxima b planetary parameters} \label{subsec:Planet_parameters}

Proxima Centauri b, has been detected using the radial velocity method. The measured revolution period is 11.186 Earth days, with the orbit semi-major axis equal to 0.0485 AU. Because of the proximity to the host star, the planet should be captured in a 3:2 or 1:1 spin-orbit resonance, depending on the orbit eccentricity $e$, as described by \cite{Ribas2016:paper}. A threshold value of $e<0.35$ was set from the observations by \cite{Anglada2016:paper}. In this work, we consider a planet captured in synchronous rotation (tidal locking) on a circular orbit ($e = 0$) with an obliquity (axial tilt) equal to zero. Such configuration causes the planet to have a permanent day-side and night-side, with a fixed sub-stellar point which always receives the same amount of stellar radiation. The principal physical and orbital parameters used in this work for Proxima b are summarized in Table \ref{Tab:Planet_parameters}. 
They indicate the exact values introduced in our models and, for this reason, the relative uncertainties of these parameters are not reported in this table.
In order to establish radius ($r_{p}$) and gravity acceleration ($g_{p}$) at the surface, we assume that the planet, for the edge-on orbital configuration, has a spherical shape with the same mean density of the Earth. The radius corresponding to the minimum mass is $r_{p} = 1.08\;R_{\oplus}$. In Section \ref{sec:Proxima_b_climate} we present a detailed climatological analysis for the edge-on configuration, thus assuming the planet minimum mass ($1.27 \; M_{\oplus}$) as its mass.
However, we perform several climate simulations with adjusted radius and mass values to correctly estimate the thermal phase curve of Proxima b.
For smaller values of the orbital plane inclination angle $\iota$, the planet increases in mass, and thus in radius. This has a significant effect on both the emission and reflected spectra of the planet.
In particular, consequences on reflected light were extensively studied by \cite{Kane2017:paper}.
To estimate the mass and radius changes for different orbital plane inclinations, we assume a simple mass-radius relation (see \citealt{Swift2012:paper,Kane2012:paper,Kane2017:paper}), and in particular the one assumed by \cite{Turbet2016:paper} for Proxima b, $r_{p} \propto (M/\sin \iota)^{0.27}$.
Although orbit inclination, true mass and radius are unknown, the planet has $\sim 84 \%$ probability to be in the terrestrial regime \citep{Kane2017:paper}.
In particular, a density transition and a planet composition dominated by volatile materials in non-terrestrial regime can be expected for a mass $> 5.1\; M_{\oplus}$, corresponding to an orbit inclination $\iota<15^{\circ}$ \citep{Kane2017:paper}.
Thus, we exclude the range $\iota<15^{\circ}$ from our analysis.
Furthermore, the recent discovery of Proxima c \citep{Damasso2020} may help in further constraining Proxima b planetary parameters. \cite{kervella2020} combine spectroscopic orbital parameters from HARPS and UVES with the astrometric proper motion anomaly (PMa) from HIPPARCOS and Gaia DR2, constraining Proxima c orbital inclination to two symmetric possible values, having the same $\sin \iota$: $152 \pm 14$ deg for a prograde orbit($90^{\circ} \leq \iota \leq 180^{\circ}$) and $28 \pm 14$ deg for a retrograde orbit($0^{\circ} \leq \iota \leq 90^{\circ}$). Assuming the coplanarity of the orbits of the two planets, the mass of Proxima b is then $2.1^{+1.9}_{-0.6} M_{\oplus}$, likely posing the planet in the terrestrial regime.

\begin{deluxetable*}{lcc}[ht!]
	\tablecaption{Proxima b Keplerian and planetary parameters assumed in PlaSim simulations.\label{Tab:Planet_parameters}}
	\tablewidth{0pt}
	\tabletypesize{\small}
	\tablehead{
	\multicolumn3c{Proxima Centauri b: derived and assumed quantities.} \\
	\colhead{Parameter} & \colhead{Symbol} & \colhead{Value}
	}
	\startdata
	Orbital period						&	$\mathrm{P}$				& 11.186 Earth days $^{\dagger}$	\\
	Orbit eccentricity					&	$e_{p}$						& 0.0 $^{\ddagger}$		\\
	Orbit semi-major axis				&	{\itshape{a}}				& 0.0485 AU $^{\dagger}$	\\
	Obliquity							&	$\alpha$					& 0.0 deg $^{\ddagger}$	\\
	Minimum mass						&	$M_{min}$			    	& $1.27 \; M_{\oplus}$ $^{\dagger}$	\\
	Mean density						&	$\rho_{p} = \rho_{\oplus}$	& $ 5514.0 \; \mathrm{kg \, m^{-3}}$ $^{\ddagger}$	\\
	Radius					     		&	$r_{p}$						& $1.08 R_{\oplus}$ $^{\ddagger}$			\\
	Surface gravitational acceleration	&	$g_{p}$						& $10.6 \; \mathrm{m \, s^{-2}}$ $^{\ddagger}$		\\
	Rotation rate						&	$\omega_{p}$				& $6.5 \times 10^{-6} \; \mathrm{rad \, s^{-1}}$ $^{\ddagger}$	\\
	\enddata
    \tablecomments{$^{\dagger}$ Measured or derived value by \cite{Anglada2016:paper}; $^{\ddagger}$ Assumed value for the simulation. Radius and surface gravitational acceleration refer to the edge-on configuration.}
\end{deluxetable*}

\section{PlaSim and libRadTran settings} \label{sec:models}

\begin{figure*}[htp]
  \centering
  \hspace*{-1cm}
  \includegraphics[width=550pt]{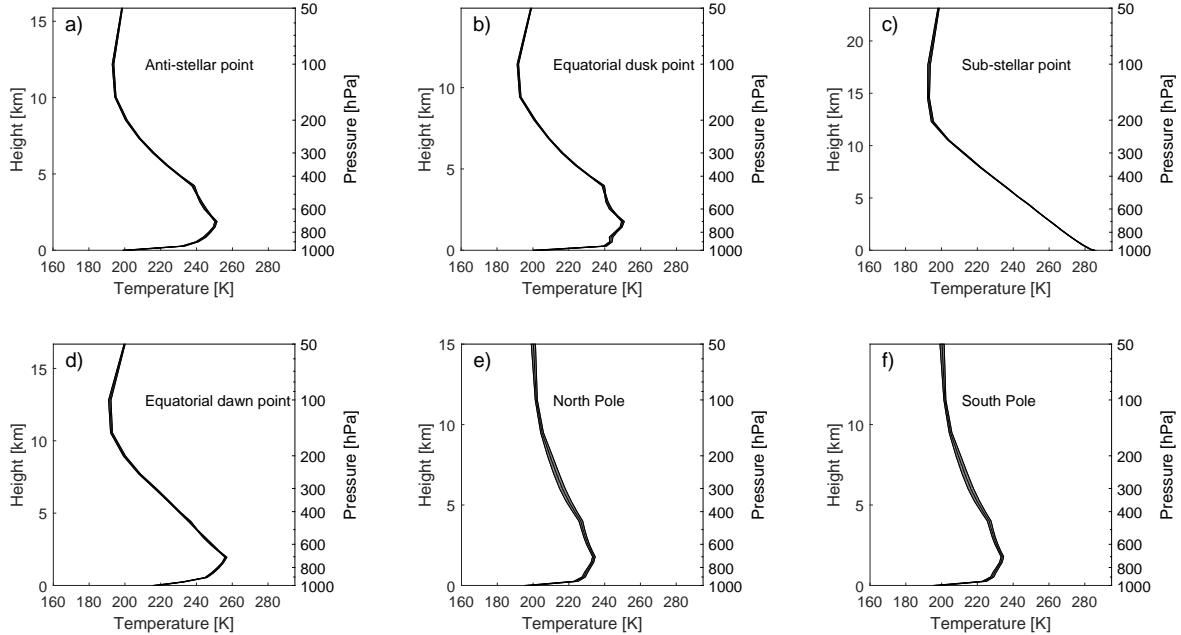}
      \caption{Temperature mean vertical profiles (black curve) and relative $\pm 1\sigma$ deviation from mean profile (gray-shaded area) for six grid-column of PlaSim simulation, relative to different regions of Proxima b: \textit{a)} anti-stellar point; \textit{b)} equator dusk point; \textit{c)} sub-stellar point; \textit{d)} equator dawn point; \textit{e)} North Pole; \textit{f)} South Pole. On the $x$-axis the value of temperature is reported, while on two $y$-axis the height and isobar level are reported, respectively. The different height of planet TOA in sub-stellar region, depends on planet thermal structure. The $1\sigma$ deviation is evaluated on the entire period of simulation (100 Earth years) and the relative maximum deviation for each panel is: \textit{a)} $\pm0.78 \; \mathrm{K}$ at surface; \textit{b)} $\pm1.32 \; \mathrm{K}$ at surface, \textit{c)} $\pm0.36 \; \mathrm{K}$ at 11.5 km; \textit{d)} $\pm0.60 \; \mathrm{K}$ at surface; \textit{e)} $\pm0.88 \; \mathrm{K}$ at surface; \textit{f)} $\pm0.87 \; \mathrm{K}$ at surface.}
      \label{Fig:Temperature_profiles}
\end{figure*}

PlaSim is a model derived from the Earth System Models (ESM) and its default configuration is set to reproduce the Earth climate. Consequently, the implemented parameterizations for the radiative transfer calculation within the atmosphere are based on the incoming solar radiation, and, in particular, on the Total Solar Irradiance (TSI) value, which is $\simeq 1361 \; \mathrm{W \, m^{-2}}$ \citep[e.g.][]{Kopp2018:paper}. Nonetheless, in order to study the atmosphere of Proxima b taking into account the different spectral energy distribution of Proxima Centauri, the parameterizations have to be modified. In particular, the stellar contribution in PlaSim simulations is parameterized by three quantities defined at the planet TOA: the total stellar irradiance $S_{0}$ in $\mathrm{W \, m^{-2}}$, the normalized fraction of $S_{0}$ in the ultraviolet and visible (UV-VIS) spectral region, $E_{1}$, and the normalized fraction of $S_{0}$ in the near infrared (NIR) spectral region, $E_{2}$. These two spectral regions are considered in our model as the shortwave radiation.
Although the output quantity provided by PlaSim are the integrated flux along all the shortwave range, the calculations in the source code are performed considering the separation between the two mentioned broadbands and the physical processes occurring within each one of them. Specifically: in the UV-VIS band, defined for wavelengths $\lambda < 0.75 \; \mathrm{\mu m}$, pure cloud scattering, ozone absorption and Rayleigh scattering are taken into account without water vapor absorption, whereas in the NIR band, defined for wavelengths $\lambda > 0.75 \; \mathrm{\mu m}$, cloud scattering and absorption and water vapor absorption are considered. Following this parameterization with a given stellar spectral irradiance $I(\lambda,T)$ in $\mathrm{W \, m^{-2} \, nm^{-1}}$, the resulting total irradiance in $\mathrm{W \, m^{-2}}$ can be obtained as
\begin{equation}
    S_{0} = \int_{0}^{\infty}I(\lambda,T)d\lambda,
\end{equation}
and consequently, the normalized energy flux fractions in the two spectral regions described above can be computed as
\begin{equation}
    E_{1} = \frac{\int_{0}^{\lambda_{\tau}}I(\lambda,T)d\lambda}{S_{0}},
\end{equation}
and
\begin{equation}
    E_{2} = 1 - E_{1},
\end{equation}
where $\lambda_{\tau} = 0.75 \; \mathrm{\mu m}$. \\
The normalized flux fractions derived from the Proxima Centauri spectrum in Figure \ref{Fig:Spectra_comparison} are $E_{1} = 0.23$, and $E_{2} = 0.77$, respectively and are modified in the PlaSim source code accordingly. \\
The longwave radiation parameterization in PlaSim includes $\mathrm{CO_{2}}$, ozone, water vapor and clouds. The scheme uses one only spectral band, with broadband emissivity and transmissivity calculations based on \cite{Manabe1961:paper} and \cite{Sasamori1968:paper} for a clear sky atmosphere and vertical discretization based on \cite{Chou2002:paper}.
This means that, as well as in the shortwave case, the radiation emitted by the planet is given as an integrated quantity along all the longwave spectral range, also known as the outgoing longwave radiation (OLR, \citealt{Petty:book}). \\
Since our aim is to study Earth-like planets, we use a simple approach and we assume an Earth-like atmosphere. The initial surface pressure is assumed equal to $1000 \; \mathrm{hPa}$.
Moreover, an ozone layer is maintained with a prescribed vertical profile parameterized by \cite{Green:1964:paper}, in order to reproduce the Earth's one, with the exception that, because of the completely different insolation pattern between Proxima b and the Earth, we neglect the typical ozone seasonality, keeping its vertical and meridional distribution constant over time. 
For an active star, as Proxima Centauri is, the impact of the stellar UV variation on the ozone vertical distribution should be considered.
Due to surface magnetic structures, the stellar UV variability can be described by [FUV-MUV] color index variation during the activity cycle (e.g., \citealt{Lovric2017:paper}, \citealt{Criscuoli2018:paper}, \citealt{Berrillietal2020:paper}).
However, for simplicity, in this work we hypothesize a constant stellar flux (i.e. no stellar variability), postponing to a future work a detailed study of the connection between stellar UV activity and planetary ozone distribution.\\
Of importance, due to its spectral properties, the albedo of sea-ice is significantly reduced around an M star with respect to a G star like the Sun (\citealt{Joshi2012:paper,Shields2013,Turbet2016:paper,Boutle2017:paper,DelGenio2019:paper} among others). In order to take this into account, in PlaSim the maximum value of broad band albedo over sea ice has been set to 0.27, based on the calculation of \cite{Turbet2016:paper}. 
In the PlaSIM code, the sea ice albedo parameterization depends on the temperature, following the formula: $R_{S} = min(R^{max}_{S}, 0.5+0.025 (273.-T_{i})$.
Where $T_{i}$ is the temperature over sea-ice and the prescribed maximum sea ice background albedo $R^{max}_{S}$, originally set to a default value of 0.7 for the Earth, has been set to 0.27 in this study.\\
The carbon dioxide concentration is maintained fixed to $360 \; \mathrm{ppm}$ during the simulation. This is chosen in order to compare our results with the works of \cite{Turbet2016:paper}, \cite{Boutle2017:paper} and \cite{DelGenio2019:paper} in which similar parameters were used.
Water vapor concentration is directly computed by PlaSim in a specific module, using pressure and temperature fields in order to determine its properties for each time-step.\\
The planet is initialized as an aquaplanet, i.e. a planet with a surface entirely covered by an ocean.
This choice is motivated by two considerations. 
Firstly, \citealt{Tian2015} showed that, in the habitable zones around M dwarfs, there are two favored types of planets with Earth-like mass which are aquaplanet and desert planets with orders of magnitude less surface water than on Earth.
Secondly, land mass distribution alters both dynamics and thermodynamics in different ways depending on the orography and the surface properties, such as albedo and soil type.
Since any land mass distribution introduced in the model would be completely arbitrary, the most straightforward choice is to totally neglect land masses.
Furthermore, although \cite{DelGenio2019:paper} showed that ocean dynamics could play a relevant role on tidally-locked planet climate, in this work we simulate a thermodynamic slab ocean of 50 meters depth, where the horizontal and vertical mixing are neglected.
The advantage of using a slab ocean instead of a full ocean (i.e. an ocean for which the 3D Navier-Stokes equations are numerically solved as well as for the atmosphere) is that it allows the climatic system to reach a steady state in less than a few decades, returning a fast numerical simulation which enables sensitivity evaluation to parameter variations. \\
In order to obtain the effect of the synchronous rotation of the planet, we modify the PlaSim radiation module source code, to fix the position of the star in the planet sky at each simulation timestep.
We run the Proxima b climate equilibrium simulation for a period of 110 Earth years with a temporal resolution of 45 minutes, that is the default timestep of the model. The output temporal resolution is 1 Earth day. Because of the spurious variability of the system introduced by the initial non-equilibrium spin-up of the simulation, a specific period at the beginning of the simulation is discarded. This is chosen following the stabilization of the surface temperature variability, i.e. when no long-term trends are found, and the mean surface temperature reach a steady state. We find that this system stabilization occurs after 10 years from the beginning of the simulation. \\
Model outputs are gridded on a $128\times64$ longitude-latitude grid, with 20 vertical pressure levels, from 1000 hPa to 50 hPa, setting the sub-stellar point above the equator, at $180^{\circ}\;\mathrm{E}$ of longitude. \\
As discussed above, PlaSim returns integrated fluxes in both shortwave and longwave spectral ranges. However, integrated fluxes are not sufficient for our purposes. 
Given that our aim is to determine whether Proxima b can be detected by analyzing its thermal emission within specific bands from current or future telescopes observations, we evaluate the planet thermal infrared emission in narrower spectral bands compared to PlaSim outputs.
In order to perform the radiative transfer calculation on the atmosphere of Proxima b, obtaining high resolution synthetic emission spectrum, we use the \textit{uvspec} model and the DISORT method. In particular, given the spectrally resolved stellar irradiance at planet TOA and the vertical profiles of the planetary atmosphere constituents, DISORT solves the 1D plane-parallel radiative transfer equation, returning radiances, irradiances, and actinic fluxes\footnote{The actinic flux is the radiant quantity used to calculate different photodissociation rates. It represents the total number of photons, or radiation, incident at a point \citep{Madronich1987:paper}.} as outputs. In this work, we use the spectral irradiance of Proxima Centauri shown in Figure \ref{Fig:Spectra_comparison} as stellar input for DISORT, whereas atmospheric vertical profiles are extracted from PlaSim simulation, considering the atmospheric columns above each grid-box. We indicate these columns as \textit{grid-columns}. Specifically, the extracted profiles within a grid-column are: \textit{i)} atmospheric pressure, \textit{ii)} temperature, \textit{iii)} air density, \textit{iv)} ozone, \textit{v)} water vapor concentration and \textit{vi)} carbon dioxide concentration. Furthermore, our implementation takes into account the cloud properties such as the liquid water content, the cloud droplet effective radius and the total optical depth. The output from \textit{uvspec} are then used to estimate the line-by-line emission spectra of the simulated planetary atmosphere, with a spectral resolution of 1 $\mathrm{nm}$, for each time-step and each grid-column from which we reconstruct thermal emission from planet TOA. Because of the imposed planetary axial and orbital symmetry, the year-to-year variability of the different atmospheric constituents (including water vapour and clouds), estimated as the standard deviation of the yearly time series over 100 Earth years, in the three regions is very small, as shown in Figure \ref{Fig:Temperature_profiles} for the temperature profile only. This small atmospheric variability allows us to assume stationary vertical profiles. This means that we can use the mean vertical profiles within each grid-column, as representative for the entire period of simulation. \\

\section{Proxima Centauri b Climate} \label{sec:Proxima_b_climate}

\subsection{Thermodynamic properties of Proxima b atmosphere} \label{subsec:Thermodynamic}

The planet surface temperature is largely used as a climatological proxy to estimate the habitability conditions on a planet (\citealt{Kasting1993:paper}) or to study the fluctuations between different multiple steady states of its climate (\citealt{Lucarini2013:paper}). 
Using the planetary parameters and atmospheric composition discussed in Section \ref{subsec:Planet_parameters}, the resulting mean surface temperature field from our PlaSim simulation shows an insolation-symmetric pattern with temperatures below the water freezing temperature $T_{ice}$
on the night-side of the planet surface, and an open ocean on most of the day-side. Surface maximum temperature is located to the east of the sub-stellar point. This is likely due to the superrotating atmosphere and will be discussed in more details later in this section.
Surface temperature ranges from a minimum value of $160 \; \mathrm{K}$ to a maximum value of $295 \; \mathrm{K}$.
These values are smaller compared to those reported by \cite{Turbet2016:paper} for similar simulation settings, that were $200 \; \mathrm{K}$ and $300 \; \mathrm{K}$, respectively, but comparable to those shown by \cite{Boutle2017:paper}. Moreover, our surface temperature maximum on the day-side is comparable with that found by \cite{DelGenio2019:paper}, whereas surface temperature minimum is about $20 \; \mathrm{K}$ higher.
Furthermore, our results regarding the planet surface region where $T > T_{ice}$ are in line with that obtained by all the three aforementioned studies for similar simulation settings.
Given the comparable results obtained using either intermediate complexity models (PlaSim in our study) or the more sophisticated GCMs (as reported in previous literature), such simplified models may represent an innovative and promising tool for future climate studies, especially to simulate future exoplanetary observations.

\begin{figure*}[htp]
  \centering
  \includegraphics[width=350pt]{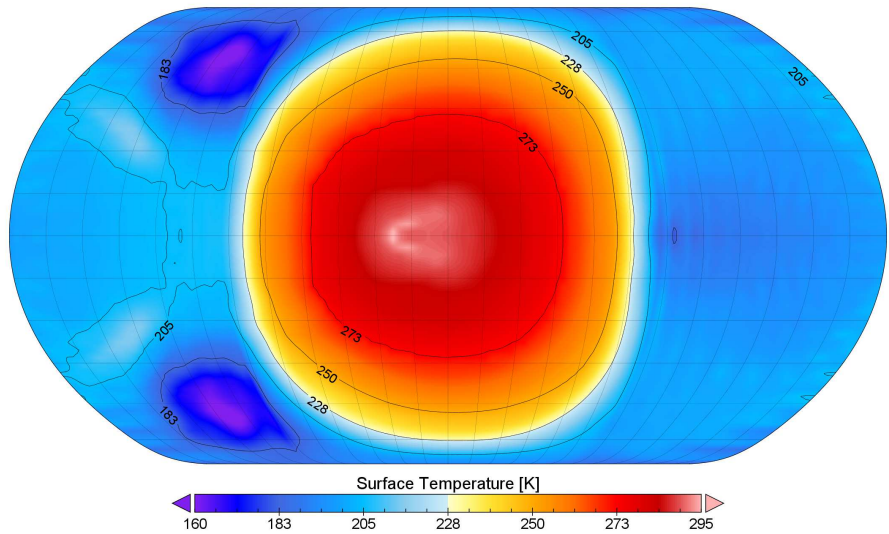}
  \caption{{\bfseries Mean surface temperature}. The tidal-locking of the planet produces the symmetrical pattern of Proxima Centauri b surface temperature: higher temperatures in the region of the sub-stellar point, with the presence of an open ocean ($T_{ice} = 273.15 K$) and colder temperatures in the night-side of the planet. Planet rotation is assumed clockwise in the simulation. This map is shown as it would be for a planet with anticlockwise (prograde) rotation, for easy comparison with previous works. See e.g. the position of the cold points in Fig.2 of \citealt{Boutle2017:paper} and Fig. 6 of \citealt{Turbet2016:paper}.}
  \label{FigSurfT}
\end{figure*}

Thermodynamic processes determine the amount of infrared radiation emitted by the planet towards space and consequently they can be used to set observational limits. One of the main absorbers in an Earth-like atmosphere is water, both as vapor and liquid (i.e. in the clouds). 
In our simulation, the atmospheric liquid water content in clouds ($w_{L}$) is lower than Earth, with a maximum of $1.73 \times 10^{-4} \; \mathrm{kg \, m^{-3}}$ located in the first layer above the planet surface in the day-side and with a dry night-side.
For reference, the maximum value of cloud liquid water content on Earth has been evaluated by \cite{Hu2007:paper} to be around $5.0 \times 10^{-4} \; \mathrm{kg \, m^{-3}}$.\\

\subsection{Atmospheric dynamics} \label{subsec:dynamics}

\begin{figure*}[htp]
  \centering
  \hspace*{-1cm}
  \includegraphics[width=550pt]{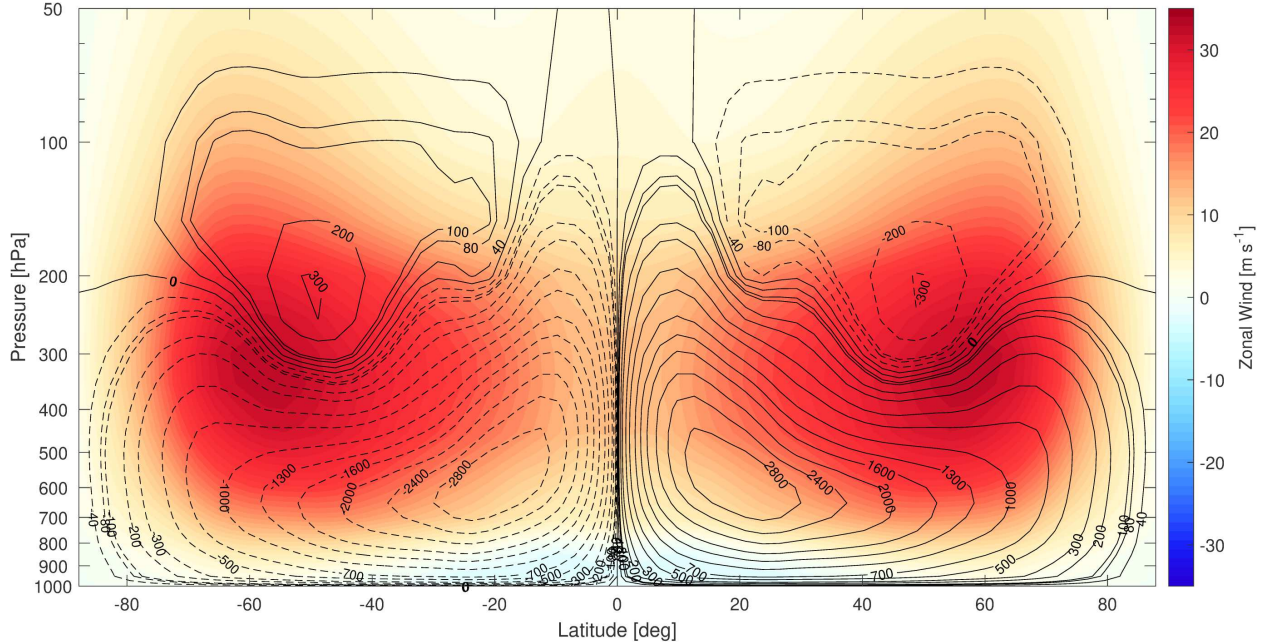}
  \caption{Zonally averaged zonal wind (colored contours) and mass stream function (black contour lines). Hadley cell-like structure in the atmosphere of Proxima Centauri b, driven by the constant insolation in the day-side. Colors represent the zonal mean of the zonal wind, where positive values are for easterlies winds (from west to the east) and negative values are for westerlies winds (from east to the west). The black lines show the mass stream function: the solid lines denote clockwise circulation ($\psi > 0$), while the dotted lines denote anticlockwise circulation ($\psi < 0$). The numbers upon the lines are the relative values of $\psi$ in $10^{6} \: \mathrm{kg \, s^{-1}}$ units. The mass stream function gives the extent of the Hadley cell-like circulation which, in the case of Proxima Centauri b, extends horizontally from the equator to poles, and vertically above 150 hPa.}
  \label{FigStreamFunction}
\end{figure*}

The planetary surface temperature pattern reported in Figure \ref{FigSurfT}, causes a convective structure in the day-side which brings the surface air masses to ascend up to 15 km. This convective flow is consistent with the planet's mean global circulation shown in Figure \ref{FigStreamFunction}. In this figure, the mass stream function $\psi$ is superimposed to the zonally averaged zonal wind. $\psi$ is defined as in \cite{Ceppi2013:paper}:
\begin{equation}
    \psi = -\frac{2\pi r_{p}}{g_{p}}\int_{0}^{p}{[\bar{v}]\cos{\theta} \, dp^{\prime}}
\end{equation}
where $g_{p}$ is the gravitational acceleration, $[\bar{v}]$ is the zonal mean of the meridional wind, $\theta$ is the latitude and $p$ is the pressure level. The mass stream function indicates the meridional flows on Proxima b and the formation of a Hadley cell-like circulation which surrounds the entire planet from equator to poles. The cell presents a clockwise branch in the northern hemisphere (solid lines in Figure \ref{FigStreamFunction}) and an anticlockwise counterpart, in the southern hemisphere (dashed lines in Figure \ref{FigStreamFunction}). This global circulation is in agreement with results from similar studies reported in the literature (e.g., \citealt{Turbet2016:paper,Boutle2017:paper,komacek2019}) for a planet rotation period of the order of one tenth of the Earth's one.

This implies that the planetary atmosphere dynamics, triggered by the constant insolation on the day-side, is driven by the interaction of the meridional circulation with the Coriolis force due to the planet rotation. This results in an equatorial symmetric jet stream at 300 hPa (Figure \ref{FigStreamFunction}). \\
Furthermore, this tidally-locked rocky planet, exhibits a superrotating atmosphere. This result is comparable to previous results (e.g., \citealt{Joshi1997:paper}, \citealt{Heng2011:paper}, \citealt{Edson2011:paper}, \citealt{Wordsworth2011:paper} \citealt{Showman2013:paper}, \citealt{Merlis2013:paper}). Superrotation is a condition whereby the atmospheric axial angular momentum is locally greater than the surface axial angular momentum. The specific equatorial angular momentum of the planet is $M_{0} = \omega r_{p}^{2}$
and it only depends on the planet radius and rotation rate, whereas the atmosphere's axial angular momentum $M_{a}$ also depends on the zonal wind and it can be expressed as
\begin{equation}
    M_{a} = r_{p}\cos\theta\left(\omega r_{p}\cos\theta + u\right),
\end{equation}
where $\theta$ is the latitude. \\
Assuming that the atmospheric angular momentum is equal to the planet equatorial angular
momentum,
\begin{equation}
    M_{a} = M_{0},
\end{equation}
a zonal wind threshold value can be derived as a function of $\theta$:
\begin{equation}
    u_{m} = \frac{\omega r_{p} \sin\theta^{2}}{\cos\theta}.
\end{equation}

\begin{figure*}[htp]
  \centering
  \hspace*{-1cm}
  \includegraphics[width=500pt]{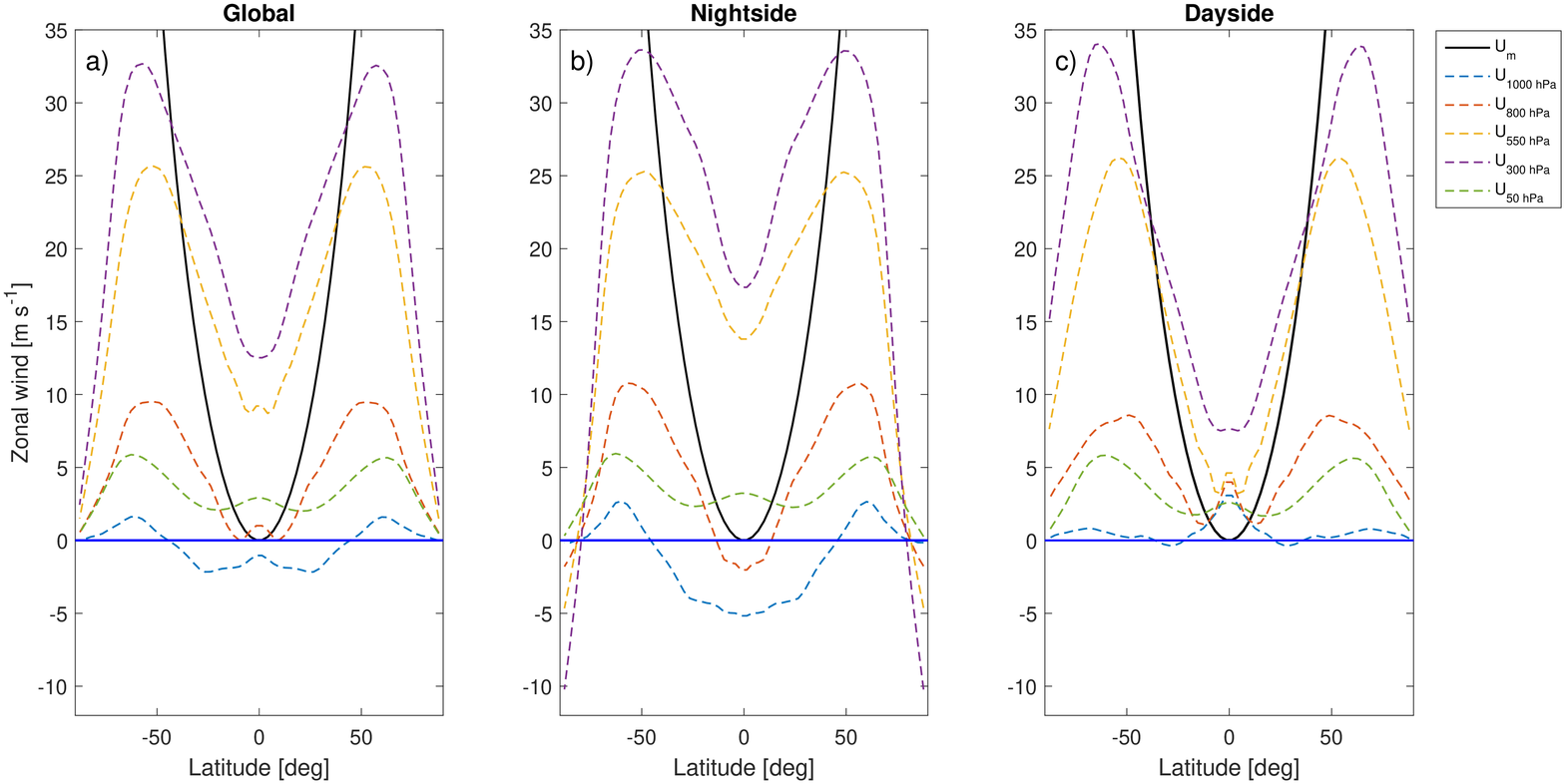}
  \caption{{\bfseries Atmospheric superrotation}. Angular momentum conserving wind $u_{m}$ (black curve) and zonal mean zonal wind (colored dashed lines) as a function of latitude $\theta$ ($x$-axis), for planet a) global mean, b) night-side mean and c) day-side mean. Each zonal mean wind curve represents the $x$ component of the wind ($y$-axis) at different pressure levels, namely $1000 \, \mathrm{hPa}$ (blue-dashed curve), $800 \, \mathrm{hPa}$ (red-dashed curve), $550 \, \mathrm{hPa}$ (yellow-dashed curve), $300 \, \mathrm{hPa}$ (purple-dashed curve) and $50 \, \mathrm{hPa}$ (green-dashed curve). The blue horizontal line represents the zero mean of the zonal wind. If a colored curve is above the solid black curve for a given latitude range, the condition $u > u_{m}$ is satisfied, and the relative atmospheric layer is in an equatorial superrotation state in that latitude range.}
  \label{FigSuprot}
\end{figure*}

Hence, $u_{m}$ represents the threshold value for which the atmosphere would be co-rotating with the planet surface at the equator. In Figure \ref{FigSuprot} the threshold value $u_{m}$ and zonal mean zonal wind at different pressure levels for Proxima b are shown for global, night-side and day-side averages. The equatorial superrotation condition is satisfied in the region between $30^\circ \, \mathrm{S} \, - \, 30^\circ \, \mathrm{N}$ with the exception of the first atmospheric layer which represents the planet surface. Such an exception seems to suggest that the aforementioned surface displacement of the warm region from the sub-stellar point is not due to superrotation. However, this result is biased by the fact that the zonal mean of zonal wind is evaluated by global averages. Nevertheless, when performing the same analysis considering the night-side and the day-side separately, the superrotation condition is satisfied also at the planet surface, but only in day-side, as shown in Figure \ref{FigSuprot}c and discussed by \cite{Showman2002:paper} and \cite{Showman2011:paper}. This result suggests that the east shift in surface temperature maximum is indeed due to the superrotating atmosphere. This also clearly demonstrates that, in case of tidally-locked planets, the average evaluations over the entire planet surface could lead to substantial misinterpretations of key climatic features and thus we strongly support the analysis of both hemispheres, separately.
Moreover, the advantage of using 3D over 1D or 2D models is evident. In fact, these latter models cannot simulate the full dynamical structure of an atmosphere, which is only possible in 3D models.

\section{Detectability of planetary atmosphere emissions from space and ground-based observations} \label{sec:detectability}

Proxima b has not yet been observed with direct imaging. This is due to the angular separation of 37 milliarcseconds between the planet and the host star \citep{Anglada2016:paper} and the planet-to-star contrast in different spectral regions. 
\cite{Lovis2017:paper} explored the possibility to observe Proxima b in the visible coupling SPHERE and ESPRESSO at VLT, to reach a contrast of $10^{-7}$. \cite{Turbet2016:paper} showed the opportunity to exploit higher planet-to-star contrast ($10^{-5}$) at 10 $\mathrm{\mu m}$ using JWST as well as the advantages that will be offered by ELT at 1 $\mathrm{\mu}$m with a contrast ($10^{-7}$) using high-resolution spectroscopy.
Observations in the thermal IR wavelengths provide distinctive information to constrain planet atmospheric properties.
However, given that the spatial resolution depends on wavelength in diffraction limited instruments, the IR direct imaging is limited to exoplanets with larger semi-major axis respect to the same technique applied in the optical spectral region.
ELT will provide a significant improvement, nevertheless, due to its diffraction limit, direct imaging for Proxima b, even for a 39 m aperture, is limited in wavelength up to 3 $\mathrm{\mu m}$ \citep{Turbet2016:paper}.
Thus, a possible strategy to overcome this limitation is to characterize the planet's atmosphere and climate studying the thermal phase curve (e.g. \citealt{Selsis2011:paper}, \citealt{Cowan2012:paper} and \citealt{Maurin2012:paper}).
We use PlaSim and the offline RTC to identify the spectral regions more sensitive to contrast variations during the orbital period. We evaluate this effect firstly for the OLR, applying the \textit{uvspec} model to PlaSim outputs. Typically, the peak of emission of an Earth-like planet in the habitable zone is expected to be in the mid-IR region around 10-20 $\mathrm{\mu m}$. 
Hence, we investigate the thermal phase curves in specific spectral bands, using \citealt{Boutle2017:paper} approach, to evaluate the planet detectability with space or ground-based instruments.

\subsection{Planetary spectra and thermal phase curves} \label{subsec:spectra}

\begin{figure*}[htp]
  \centering
  \hspace*{-1cm}
  \includegraphics[width=540pt]{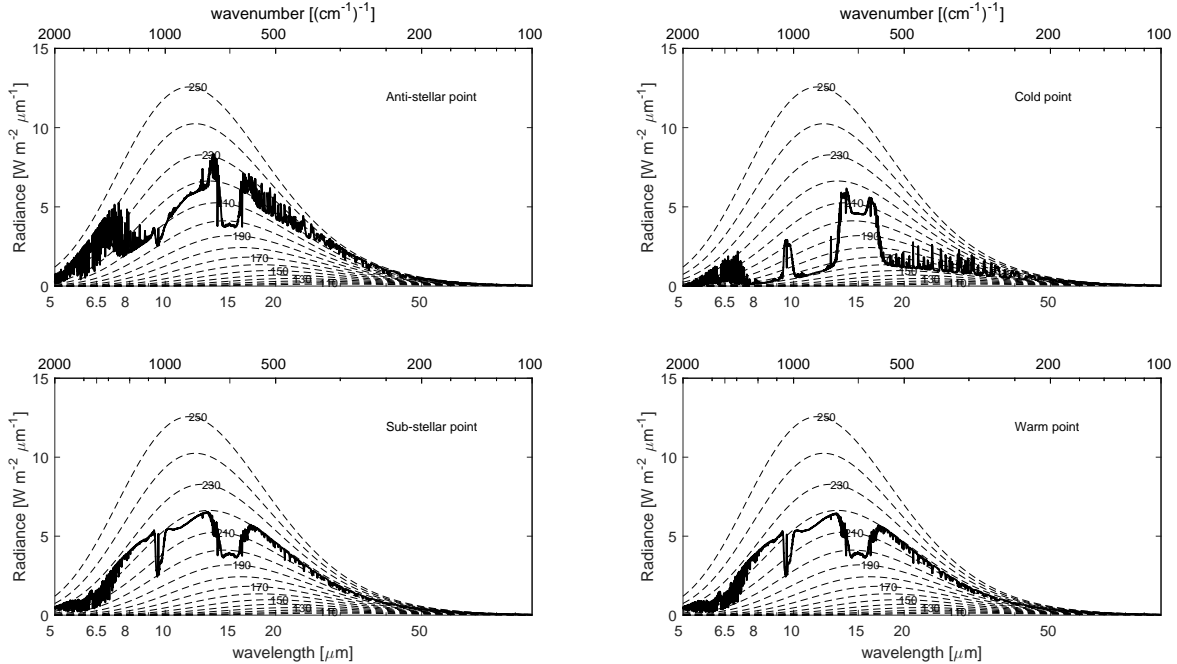}
  \caption{{\bfseries Infrared emission spectra}. Infrared emission spectra at Proxima b TOA, as a function of wavelength $\lambda$ (bottom x-axis) and wavenumber $\tilde{\nu}$ (top x-axis) are obtained with {\itshape uvspec}, above different regions of the planet: 
 a) the anti-stellar point ($0^{\circ}\mathrm{N} \, , \, 0^{\circ}\mathrm{E}$); b) the  cold point ($66^{\circ}\mathrm{N} \, , \, 290^{\circ}\mathrm{E}$); c) the sub-stellar point ($0^{\circ}\mathrm{N} \, , \, 180^{\circ}\mathrm{E}$) and d) the warm point ($1^{\circ}\mathrm{N} \, , \, 199^{\circ}\mathrm{E}$). Dashed lines represent the ideal black body emission from Planck's law curves relative to the black body temperatures from 100 K to 300 K, with 10 K separation. The radiative effect of the atmospheric component can be seen on the curves and in particular, the carbon dioxide signature, around $\lambda =$ 15 $\mathrm{\mu m}$ ($\tilde{\nu} =$ 666 $\mathrm{cm^{-1}}$), the ozone signature, around $\lambda =$ 9.6 $\mathrm{\mu m}$ ($\tilde{\nu} =$ 1040 $\mathrm{cm^{-1}}$) and the water vapor signatures at $\lambda < $ 7 $\mathrm{\mu m}$ ($\tilde{\nu} >$ 1300 $\mathrm{cm^{-1}}$).}
  \label{FigIRSpectra_TOA}
\end{figure*}

To evaluate spectral properties of the planetary atmosphere, we follow the method described in Section \ref{sec:models}. Planetary spectral thermal fluxes at the TOA for different planet regions, are shown in Figure \ref{FigIRSpectra_TOA}. Because of the synchronous rotation, the planet thermal flux has a strong dependence from longitude and latitude, with the peak of emission at $\lambda \approx 14 \; \mathrm{\mu m}$. The \textit{uvspec} module applied to the PlaSim outputs returns 8192 thermal spectra, one for each latitude/longitude grid-column of the simulation (i.e. $64\times128$ spectra). In Figure \ref{FigIRSpectra_TOA}, only four of these spectra are shown as reference: the thermal spectra as seen from the planet TOA above the sub-stellar and anti-stellar points and above
the warm and cold points.
The carbon dioxide absorption/emission feature can be seen at $\lambda\approx 15 \; \mathrm{\mu m}$ (i.e. $\tilde{\nu}\approx 600 \; \mathrm{cm^{-1}}$) in all panels and it is clear how its emission temperature remains the same in each region, being $\sim \; 200 \, \mathrm{K}$. In particular, on the planet night-side, the carbon dioxide signature is clearly visible being the feature with the maximum peak of emission.

\begin{figure*}[htp]
  \centering
  \includegraphics[width=300pt]{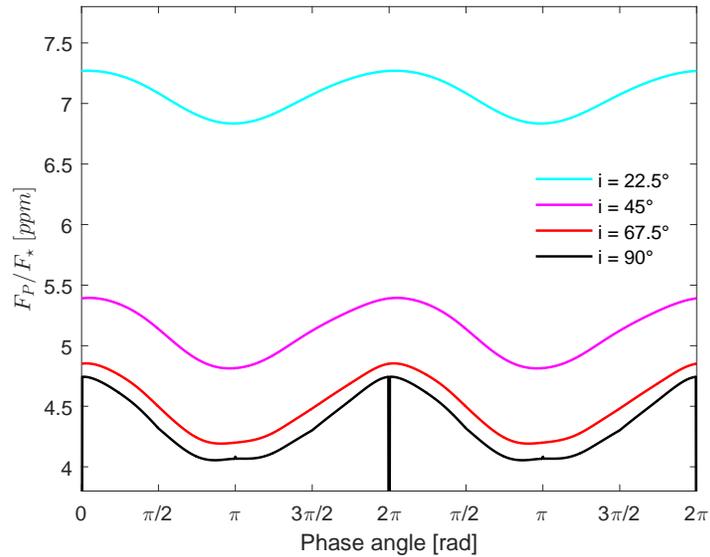}
  \caption{{\bfseries Planet phase curves.} Planet/Star contrast in the OLR region ($3 \; \mathrm{\mu m} \leq \lambda \leq 100 \; \mathrm{\mu m}$), during two complete orbits of Proxima b around the host star. Each curve represents a different inclination angle of the planet orbital plane with respect to the observer. The black curve represents the thermal emission for an edge-on geometry, i.e. the case of a transiting planet. For this reason, the value of $\Phi(\xi,90)$ drops to zero at $\xi = 2\pi$ (secondary eclipse). Also, the small effect due to primary eclipse is visible at $\xi = \pi$.}
  \label{Fig_phase_curve}
\end{figure*}

The planet emission spectra is used to derive the thermal phase curve.
The expected flux from Proxima b is obtained integrating the thermal emission over the whole surface of the planet taking into account the distance and the position angle. Since the exact geometry of the planet-star system is not currently known we consider several possible inclinations of the orbital plane with respect to an observer on Earth. 
For this purpose, we use the visibility function
\begin{eqnarray}
    V(\theta,\phi,t,\theta_{0}) = \nonumber \\
    \max\{\sin(\theta)\sin(\theta_{0})\cos \left[\phi-\phi_{0}(t) \right]+ \nonumber \\
    \cos(\theta)\cos(\theta_{0}), 0\}
\end{eqnarray}
described by \cite{Cowan2013:paper}, to obtain the visible region of the planet as a function of the orbital parameters at given orbital time.
\begin{figure*}[htp]
  \centering
  \includegraphics[width=450pt]{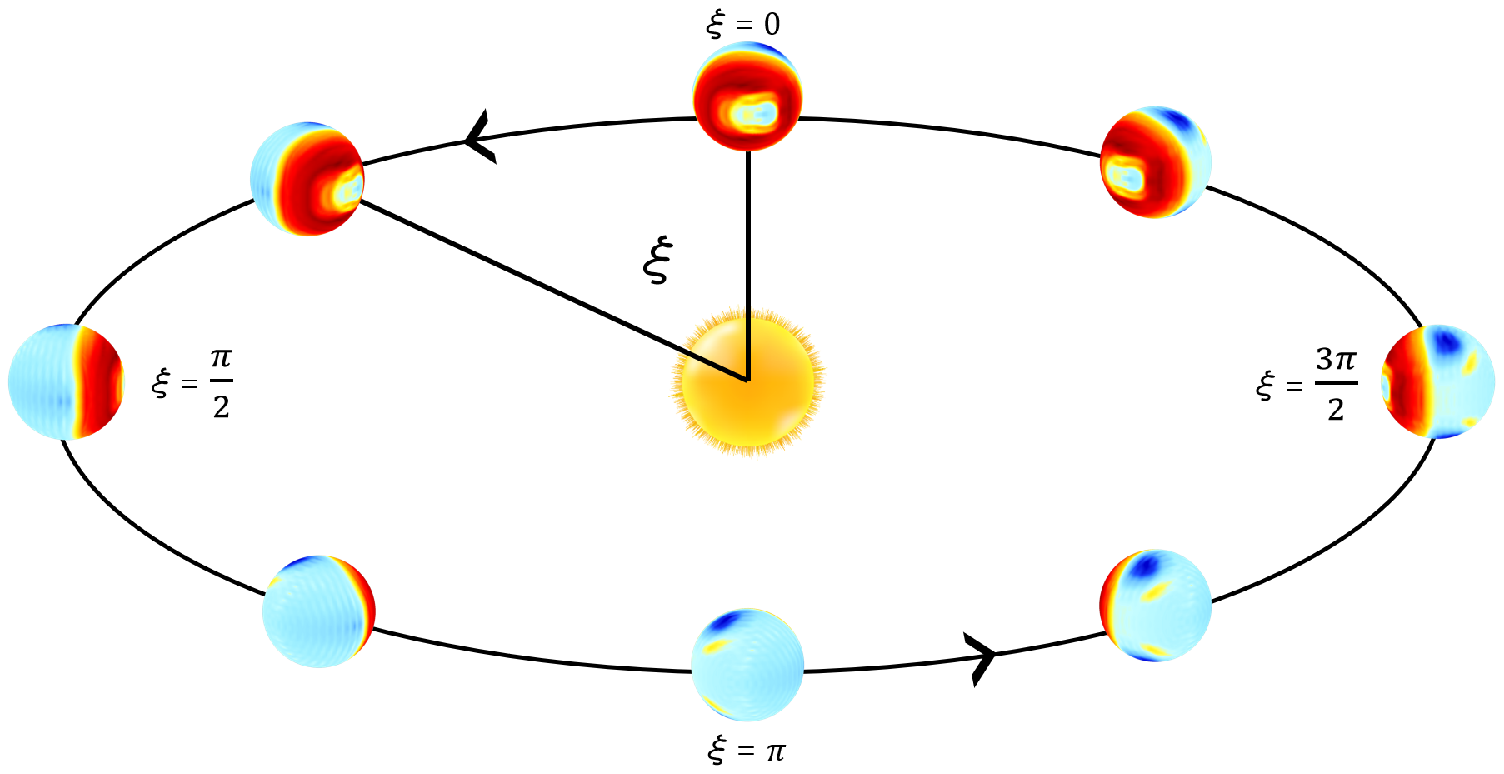}
  \includegraphics[width=300pt]{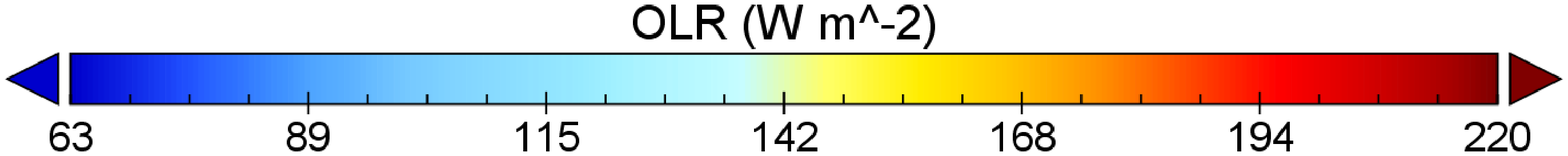}
  \caption{{\bfseries Proxima b orbit.} The figure represents Proxima b on its circular orbit around the host star. Because of the 1:1 spin/orbit resonance, an outside observer can explore all sides of the planet in a single revolution. Surface color represents the amount of OLR emission: red regions represent the emission maximum, whereas the dark blue regions represent the emission minimum.}
\end{figure*}
The visibility function $V$, depends on 4 parameters: \textit{i)} the planet longitude $\phi \in (0, 2\pi]$; \textit{ii)} the planet co-latitude $\theta \in (0, \pi)$;
\textit{iii)} the sub-observer longitude $\phi_{0} \in (0, 2\pi]$; and \textit{iv)} the sub-observer co-latitude $\theta_{0} \in (0, \pi)$. \\
The sub-observer longitude is a function of the orbital time $t$ and it can be linked to the phase angle $\xi$ (see Figure \ref{Fig_phase_curve}a), via the relationship $\xi = \Omega t,$ where $\Omega$ is the planet orbital angular velocity (i.e. the revolution rate). Since we assumed a tidally-locked planet, the revolution period is equal to the rotation period and $\Omega = \omega$. The sub-observer co-latitude instead, corresponds to the planet orbital plane inclination $\iota$, with respect to the observer. \\
Finally, by performing the substitution $\theta_{0}\rightarrow\iota$, the planet thermal emission $F_{p}(\xi,\iota)$ is given by:
\begin{equation}
    F_{p}(\xi,\iota) = \int_{0}^{2 \pi}\int_{0}^{\pi} I(\theta,\phi)V(\theta,\phi,\xi,\iota) d\theta d\phi,
\end{equation}
where $I(\theta,\phi)$ is the planet integrated thermal flux in a selected spectral range $\lambda_{1} \leq \lambda \leq \lambda_{2}$, for each latitude/longitude box of the simulation. \\

The integrated flux from Proxima Centauri, $F_{\star}$, is considered constant with time for any wavelength range. This means that we are neglecting any possible influence of the stellar variability. We define the planet/star contrast $\Phi(\xi,\iota)$:
\begin{equation}
	\Phi(\xi,\iota) = \frac{F_{p}(\xi,\iota)}{F_{\star}},
\end{equation}
where $F_{p}$ is the planet atmospheric emission. Both $F_{p}$ and $F_{\star}$ are evaluated in the same spectral range. In particular, we are interested in the amplitude $A(\iota)$ of the planet thermal phase curve defined as
\begin{equation}
    A(\iota) \equiv \Phi_{max}(\xi,\iota)-\Phi_{min}(\xi,\iota),
\end{equation}
since the amplitude modulation of the phase curve is the measurable quantity for unresolved exoplanetary systems.
Furthermore, $A(\iota)$ must be compared to the achievable photometric precision, which crucially depends on the photons available in the considered spectral range. The ratio between these two quantities identifies the most suitable band for the observations.

In Figure \ref{Fig_phase_curve}b, the planet thermal phase curve computed in the OLR spectral range ($3 \; \mathrm{\mu m} \leq \lambda \leq 100 \; \mathrm{\mu m}$) is shown for different orbital plane inclinations.

Although we evaluated the thermal phase curves for several inclination angles $0^{\circ} \leq \iota \leq 180^{\circ}$, here we show only the results obtained for $\iota \leq 90^{\circ}$, because the system symmetry produces completely specular curves for $\iota > 90^{\circ}$. 
The presence of atmosphere and its role in heat advection produces a not perfectly symmetric phase curve.\\ 
\vspace*{1cm}

\subsection{Proxima b detectability from space observatories} \label{subsec:space_observations}

Future detection of Proxima b thermal emission is evaluated considering the technical specification of the James Webb Space Telescope (JWST), specifically the case of the imaging mode of the Mid-Infrared Instrument (MIRI, \citealt{Wells2015:paper}). The photometric bands of the MIRI imager (MIRIM, \citealt{Bouchet2015:paper}) are considered and we compute the amplitude of the phase curve, fixing the orbital inclination angle to $90^{\circ}$.
Given the Proxima Centauri spectrum, the photometric precision achievable in each band is also computed, expressed in terms of the minimum detectable amplitude $A_{0}^{(N)}$. This is calculated assuming photon noise limited observations and it will be extensively described later in this section.
In Table \ref{tab:filter_choise} these quantities are compared and the ratio between the two amplitudes is found to be maximum in the MIRI F2100W passband filter, corresponding to the spectral band $18.5 \; \mathrm{\mu m} \leq \lambda \leq \mathrm{23.5 \; \mu m}$.

\begin{deluxetable*}{lccc}[!ht]
\tablecaption{Amplitude of the phase curve $A(90^{\circ})$, minimum detectable amplitude $A_{0}^{(N)}$ (24 hours integration time) and ratio between them for the photometric bands of JWST/MIRIM.
\label{tab:filter_choise}}
\tablewidth{0pt}
\tablehead{
\colhead{MIRIM filter} & \colhead{$A(90^{\circ}) \; (ppm)$} & \colhead{$A_{0}^{(N)} \; (ppm)$} & \colhead{$A(90^{\circ})/A_{0}^{(N)}$}
}
\startdata
F560W  & 0.06 & 1.6 & 0.04	\\
F770W  & 2.4 & 1.7 & 1.4	\\
F1000W  & 8.3 & 2.4 & 3.5	\\
F1130W  & 15.0 & 5.2 & 2.9	\\
F1280W  & 13.6 & 3.4 & 4.0	\\
F1500W  & 4.0 & 3.8 & 1.1	\\
F1800W  & 24.7 & 4.9 & 5.0	\\
F2100W  & 34.3 & 4.7 & 7.3	\\
F2550W  & 14.0 & 9.9 & 1.4	\\
\enddata
\end{deluxetable*}

In Table \ref{tab:amplitude} the maximum values of $\Phi(\xi,\iota)$, $\Phi_{max}$ and the relative amplitude $A$, evaluated in the above mentioned spectral band, are reported as a function of the system inclination.
The geometric probability of transit for Proxima b is very small ($\sim 1.5 \%$, \citealt{Anglada2016:paper}) and recent Spitzer Space Telescope observations ruled out planetary transits at the 200 ppm level at $\mathrm{4.5 \mu m}$ \citep{no-transit6}. Thus, the effect of a transit is excluded from subsequent analysis. The $\iota = 90^{\circ}$ case has to be intended as an almost edge-on configuration with no-transit.
The amplitude $A$ is expected to decrease with inclination, considering that day-side becomes partially visible along the full orbit. Nevertheless, this decrease is compensated by the increase of the planet mass and, in turns, its radius (\citealt{Turbet2016:paper}). The amplitude $A$ results in this case almost constant with respect to the orbital inclination. Our results are also in line with \cite{Kane2017:paper} results regarding the planet/star contrast.

\begin{deluxetable*}{lcc}[!ht]
\tablecaption{
Planet/star thermal emission maximum contrast $\Phi_{max}$ and amplitude $A$ for different orbital inclination angle $\iota$ (MIRI imager F2100W).
\label{tab:amplitude}}
\tablewidth{0pt}
\tablehead{
\colhead{$\iota$} & \colhead{$\Phi_{max} \; (ppm)$} & \colhead{$A \; (ppm)$}
}
\startdata
$22.5^{\circ}$  & 364 & 21.1	\\
$45^{\circ}$    & 271 & 23.4    \\
$67.5^{\circ}$  & 245 & 32.7	\\
$90^{\circ}$    & 240 & 34.3 	\\
\enddata
\end{deluxetable*}

In order to derive the detectability limits of Proxima b using the JWST, the signal-to-noise ratio on the MIRIM detector is computed for the passband filter F2100W. Assuming a photon noise limited observation, two different methods are used to evaluate the photometric precision. 
In both cases we consider an exposure time $\tau = 5$ hours, which is a small fraction of the planet orbital period ($\tau^{*} < P/50$).
The observing strategy simulated with the JWST Exposure Time Calculator (ETC)\footnote{\url{https://jwst.etc.stsci.edu/}} consists of several short exposures ($\sim 1.2$ s) to avoid detector saturation. The ETC simulation was set to use the SUB128 subarray of the MIRI detector in fast mode. Proxima Centauri is a bright source, so an individual exposure of 1.2 s almost exploits the full dynamic range of the camera, thus the readout noise is not a limiting factor for the observations. The readout time for the SUB128 subarry is 0.119 s, therefore the duty cycle is $\sim$ 90\% and 5.5 hours are needed for a 5 hours exposure time.

Our first method consists in computing the number of photons, within the assigned spectral band, collected by the instrument, as per \cite{Yang2013:paper}:
\begin{equation}
    N = \varepsilon\frac{\pi\mathcal{A}_{T}^{2}}{4}\tau\left(\frac{R_{\star}}{D}\right)^{2}\int_{\lambda_{1}}^{\lambda_{2}}\frac{F_{\star}(\lambda)}{E(\lambda)}d\lambda,
\end{equation}
where $\mathcal{A}_{T}$ is the telescope collecting area, $\tau$ is the exposure time, $R_{\star}$ is the star radius, $D$ is the star-observer distance, $F_{\star}(\lambda)$ is the stellar spectral energy distribution and $E = hc/\lambda$ is the energy of a photon for a given wavelength $\lambda$. 
In the case of JWST, the primary mirror effective collecting area is  $\mathcal{A}_{T}= 25 \; \mathrm{m^{2}}$ (\citealt{Gardner2006:paper}).
Furthermore, we also consider the factor $\varepsilon$ in order to take the MIRIM filter efficiency into account, which is 25\% for the F2100W filter \footnote{\url{https://jwst-docs.stsci.edu/display/JTI/MIRI+Filters+and+Dispersers}}. 

\begin{figure*}[htp]
  \centering
  \includegraphics[width=250pt]{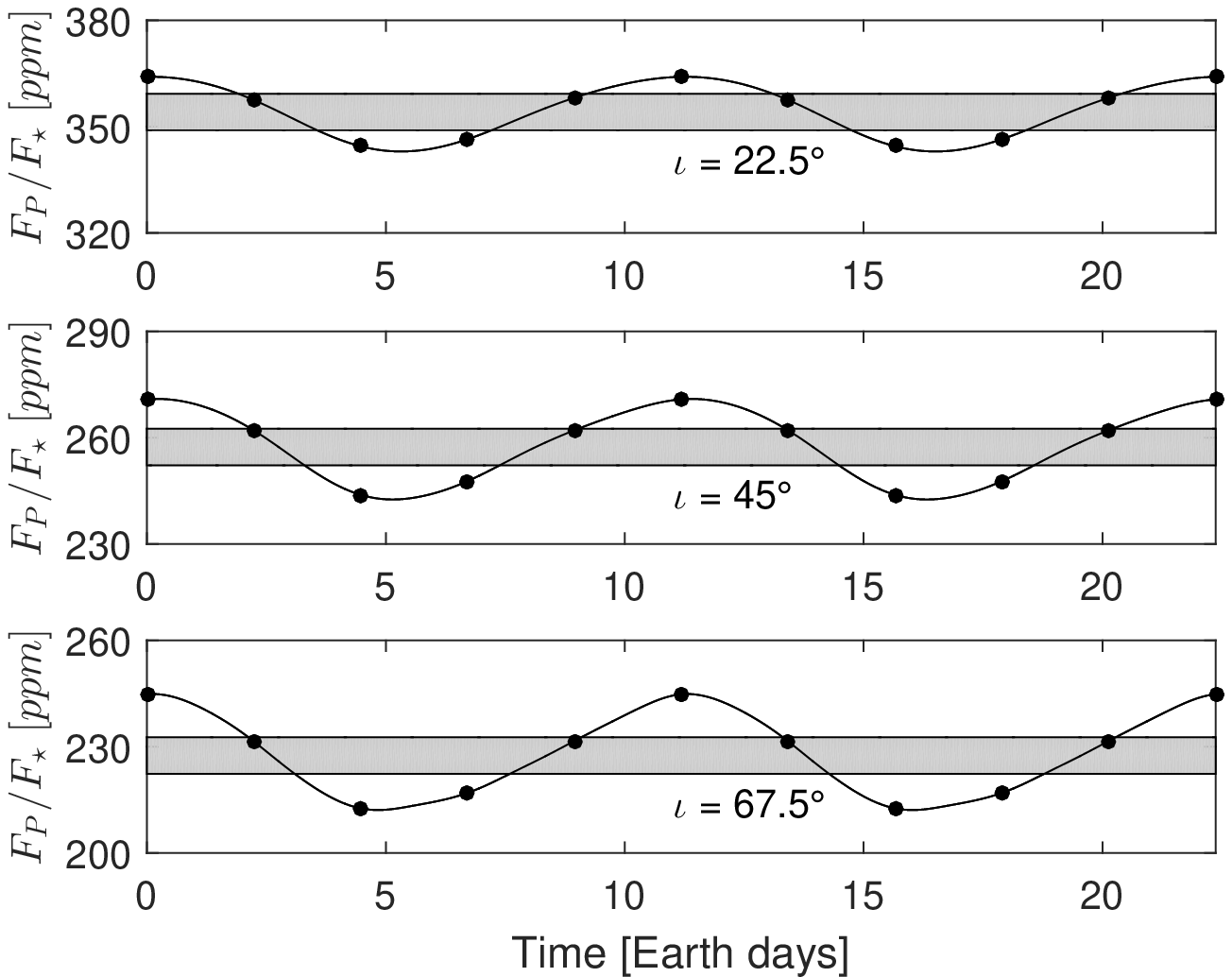}
  \includegraphics[width=250pt]{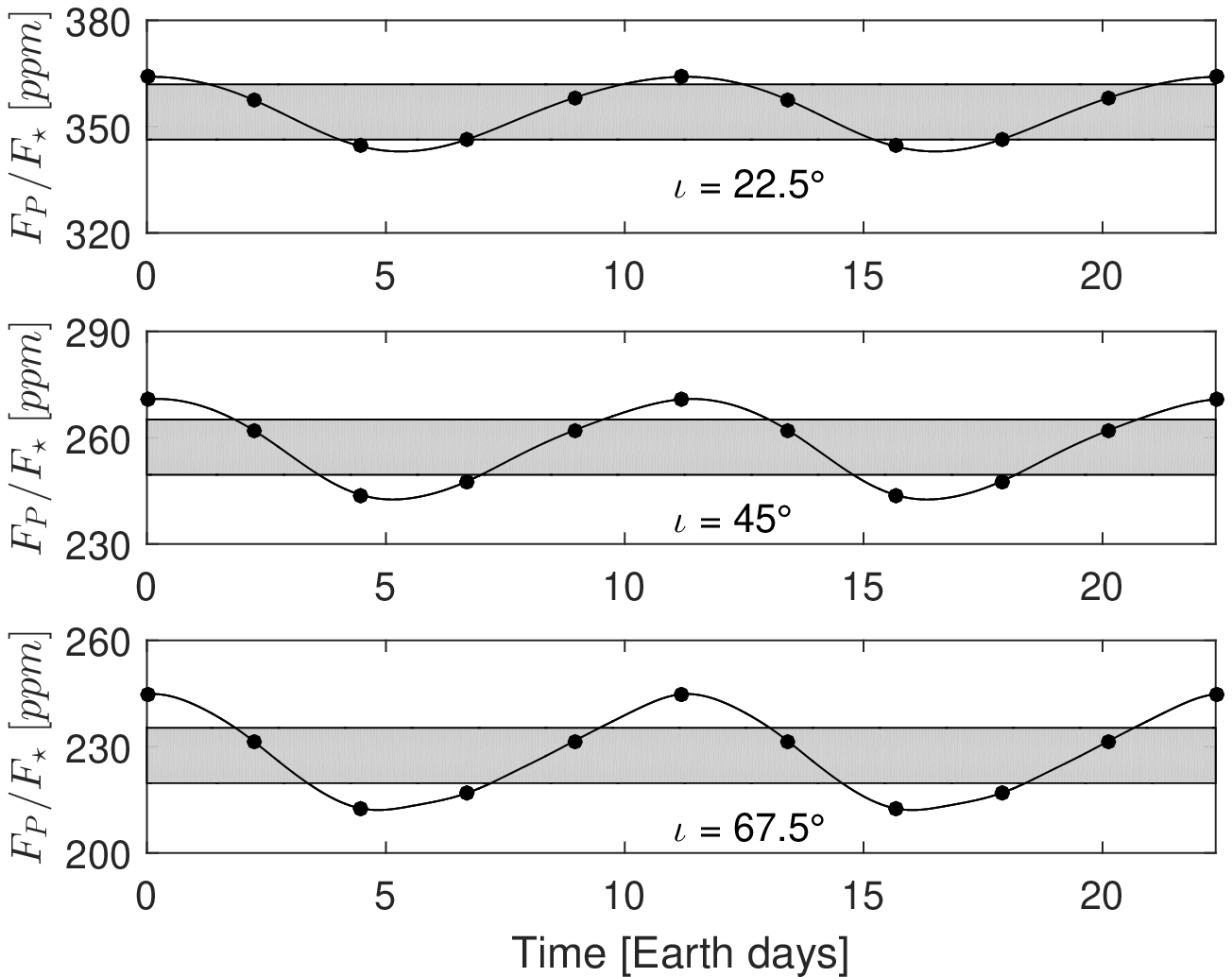}
  \caption{{\bfseries Five hours exposure time detection limits.} The planet/star contrast (y-axis) in the spectral band $18.5 \div 23.5 \; \mathrm{\mu m}$ as a function of time for a sample of three possible planet orbital plane inclination angles. Gray area represent the minimum detectable amplitude of the phase curve at 1-$\sigma$ level for Proxima b (5 hours exposure time) obtained considering the photon number: a) derived from the stellar flux following \cite{Yang2013:paper} (left) and b) the Exposure Time Calculator (ETC) of JWST, with MIRIM settings (right).
  }
  \label{Fig_flux_variation}
\end{figure*}

The instrument photometric precision is estimated considering a photon shot noise limited observation, obtaining $PP_{N} = 5.2 \; \mathrm{ppm}$.
The corresponding minimum detectable amplitude of the phase curve at 1-$\sigma$ level is equal to twice the computed photometric precision: $A_{0}^{(N)} = 10.3 \; \mathrm{ppm}$.
In Figure \ref{Fig_flux_variation}a, the gray-shaded bands have an amplitude equal to the computed $A_{0}^{(N)}$.

The second method we use to evaluate the photometric precision of JWST exploits the ETC.
The JWST ETC estimates the infrared background using a model including many celestial sources, \textit{e.g.}, zodiacal light, interstellar medium, and cosmic infrared background \citep{Reach1997, kelsall1998, Krick2012}, in addition to telescope thermal and scattered light. Moreover, it achieves accurate calculation of signal-to-noise ratio, considering the MIRIM instrumental characteristics and the filter response functions. Using JWST ETC, we obtain a photometric precision $PP_{ETC} = 7.8 \; \mathrm{ppm}$, and a corresponding $A_{0}^{(ETC)} = 15.6 \; \mathrm{ppm}$ (see Figure \ref{Fig_flux_variation}b).
This means that the photometric performance evaluated with the ETC, is lower but comparable with the one obtained using the method of \cite{Yang2013:paper}.

\begin{figure*}[htp]
  \centering
  \includegraphics[width=250pt]{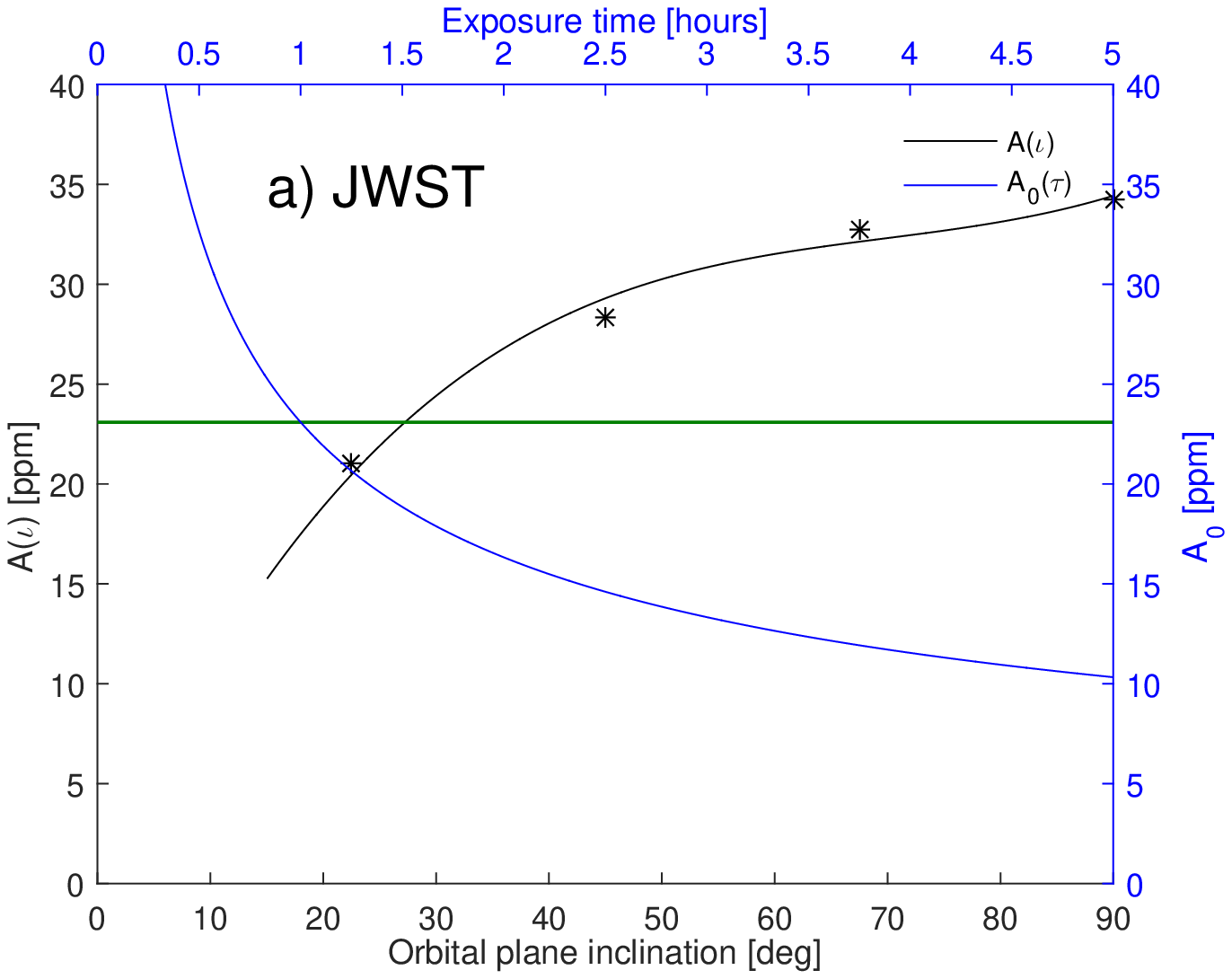}
  \includegraphics[width=250pt]{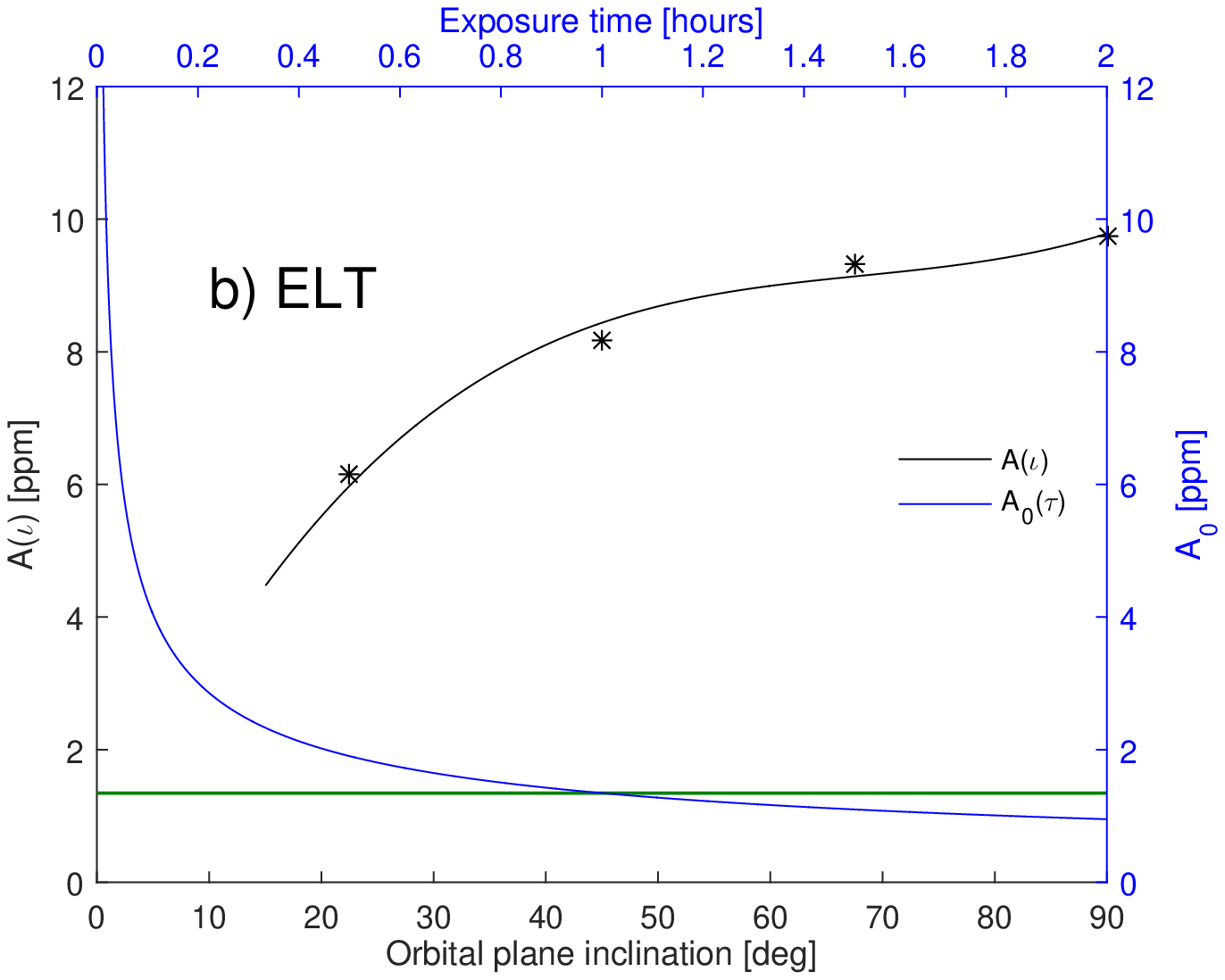}
  \caption{{\bfseries Proxima b atmosphere detection comparison.} Comparison between JWST-MIRIM ($18.5 \div 23.5 \; \mathrm{\mu m}$), panel a, and ELT-METIS ($8.3 \div 13.8 \; \mathrm{\mu m}$), panel b, for $1\sigma$ detection of broadband Proxima b atmospheric emission. The black curves represent the semi-empirical thermal phase curve amplitude $A$ as a function of the planet orbital plane inclination angle $\iota$. Values of this function below $\iota = 15^{\circ}$ are not shown since they are outside the validity range of our model. Star symbols represent sample values of the amplitude $A$ for the fixed inclination angles reported in Table \ref{tab:amplitude}. The blue curves represent the minimum detectable amplitude $A_{0}$ as a function of exposure time, evaluated as in \cite{Yang2013:paper}. The green horizontal line indicates the values of $A_{0}$ relative to an exposure time of 1 hour in both panels.}
  \label{Fig_A0_i_pp_tau}
\end{figure*}

In Figure \ref{Fig_A0_i_pp_tau} we combine the instrumental photometric performance and the planet thermal phase curve amplitude together in order to derive the lower limit of the exposure time given a certain planet orbital plane inclination angle. In particular, we compare $A_{0}^{(N)}$ as a function of $\tau$ and $A$ as a function of $\iota$.
A semi-empirical equation for $A(\iota)$ is obtained by fitting with a cubic function the four points resulting from our simulation and reported in Table \ref{tab:amplitude}.
From this figure, we can also establish for which exposure time the instrumental photometric precision $A_{0}^{(N)}$ is sufficient for the detection of signature of the planet of amplitude $A(\iota^{\ast})$, given its dependency on the inclination.
For instance, in the case shown in Figure \ref{Fig_A0_i_pp_tau}a, an exposure time of $\sim$ 5 hours is necessary to detect planet thermal emission for all considered inclinations.
We recall that, assuming coplanarity of Proxima b and Proxima c as in \citealt{kervella2020}, it is possible to constrain the orbital plane inclination angle between 14$^{\circ}$ and 42$^{\circ}$. The amplitude of the phase curve for those inclination is expected to be between $\sim$ 15 ppm and $\sim$ 30 ppm in the band $18.5 \div 23.5 \; \mathrm{\mu m}$.\\

\subsection{Proxima b detectability from ground-based observatories} \label{subsec:ground_observations}

In the previous section, the JWST photometric accuracy in the case of Proxima b observations has been computed. Such an approach can in principle be applied to similar observations performed from ground-based observatories. However, for ground-based infrared observations, mainly at mid-IR wavelengths \citep[e.g.][]{Berrillietal1987:paper, Berrillietal1989:paper, Berrillietal1992:paper}, the background is large when compared to the emission of stellar sources. Further, this background emission is very sensitive to the telescope thermal and scattered light (at 10 microns the telescope emissivity can dominate over sky emission) and atmospheric status can fluctuate rather fast in time and in space.\\
Here, we compare estimated $1\sigma$ detection for both JWST-MIRIM and the mid-infrared imager and spectrograph METIS (\citealt{Brandl2014}) which will be installed at the Extremely Large Telescope (ELT) (\citealt{Gilmozzi2007:paper}). ELT will be the largest optical/NIR telescope in the world with a primary mirror of 39 meters diameter. The technical characteristics of the ELT METIS needed to perform our evaluation have been obtained from \cite{Brandl2014}, \cite{Brandl2016}, and \cite{Brandl2018}. An attempt to estimate the performance of ELT METIS through a sensitivity model introducing a variety of sources is presented in \cite{Kendrew2010}. However, in this work, we do not make assumptions about the noise introduced by the instrument and the Earth atmosphere by limiting the analysis to only photometric shot noise.
In order to estimate the photometric precision of METIS we assume a primary mirror collecting area for ELT $\mathcal{A}_{ELT} = 978 \; \mathrm{m^{2}}$ and a 10\% efficiency.

\begin{deluxetable*}{lcccccc}[htp]
\tablecaption{Effective wavelengths, full width half maxima, zero-magnitude fluxes for the ESO infrared passbands used in this work, taken from \citealt{vdBliek:paper}. We also report the corresponding amplitude of the phase curve $A(90^{\circ})$, minimum detectable amplitude $A_{0}^{(N)}$ (1 hour integration time) and ratio between them for the same photometric bands. \label{tab:zeromagnitude}}
\tablewidth{0pt}
\tablehead{
\colhead{Filter} & \colhead{$\lambda_{eff.}$} & \colhead{FWHM} & \colhead{$F_{\lambda}(m_{\lambda}=0)$} & \colhead{$A(90^{\circ})$} & \colhead{$A_{0}^{(N)} $} & \colhead{$A(90^{\circ})/A_{0}^{(N)}$} \\
\colhead{} & \colhead{$\mathrm{\mu m}$} & \colhead{$\mathrm{\mu m}$} & \colhead{$\mathrm{W m^{-2} nm^{-1}}$} & \colhead{ppm} & \colhead{ppm}
}
\startdata
L'      & 3.771  & 0.580 & $5.58 \times 10^{-14}$ & 0.007 & 1.9 & 0.004 \\
M       & 4.772  & 0.381 & $2.21 \times 10^{-14}$ & 0.065 & 2.3 & 0.03 \\
N       & 11.055 & 5.47  & $1,29 \times 10^{-15}$ & 9.7 & 1.7 & 16.5 \\
$Q_{0}$ & 18.666 & 3.23  & $1,09 \times 10^{-16}$ & 32.7 & 5.0 & 6.54 \\
\enddata
\end{deluxetable*}

The amplitude $A(\iota)$ is computed for the N band ($8.3 \div 13.8 \; \mathrm{\mu m}$, see also Table \ref{tab:zeromagnitude}) and it is shown as black star symbols in Figure \ref{Fig_A0_i_pp_tau}.
The values of $A(\iota)$ are smaller compared to those evaluated by JWST-MIRIM, this is mainly due to the different spectral intervals used. 
Considering a 1 hour exposure time for ELT-METIS, a minimum detectable amplitude $A_0 = 1.7 \; \mathrm{ppm}$ is computed, which would be sufficient to detect planetary signature for all the range of considered orbital inclinations. Given the noise assumptions, the computed exposure time (1 hour) must be considered as a lower limit to achieve such a photometric precision.
Given that the ELT aperture is considerably bigger compared to the JWST one and only considering photon shot noise, the ELT photometric precision in the N band is almost 14 times smaller than the JWST in the F2100W band.

\begin{figure*}[htp]
  \centering
  \hspace*{-1cm}
  \includegraphics[width=500pt]{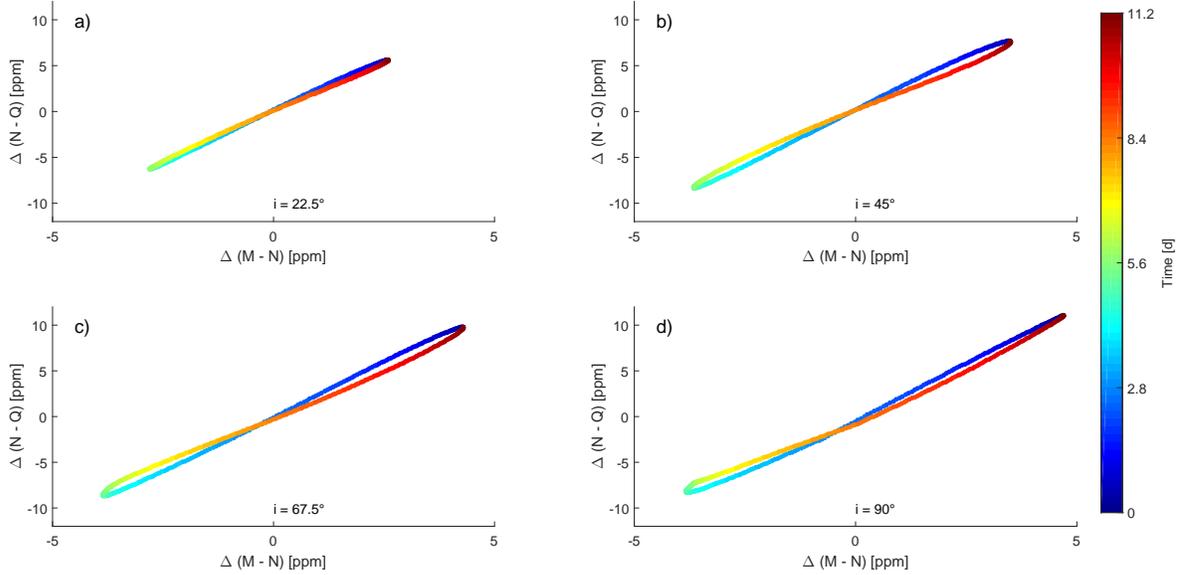}
  \caption{{\bfseries Color-color scatter plot}. The color variation of the Proxima Centauri/Planet b system during one full orbit for four different orbital plane inclination angles, respectively $22.5^{\circ}$, $45^{\circ}$, $67.5^{\circ}$, and $90^{\circ}$. The color variation is obtained evaluating the magnitude of the system in three different bands reported in Table \ref{tab:zeromagnitude}.}
  \label{Fig_color-color}
\end{figure*}

Finally, a general method to detect and characterize exoplanets using the information coming from different IR bands is introduced. For this reason, the variation in time of the color of exoplanetary systems are analyzed. Calculations are performed using the ESO IR photometric bands, indicated by the letters $\mathbf{M}$, $\mathbf{N}$ and $\mathbf{Q_{0}}$, to maximize the planet contribution. These bands, with the relative zeropoint magnitude fluxes, are described in \cite{vdBliek:paper} and summarized in Table \ref{tab:zeromagnitude}. We compute the color indices $\mathbf{M} - \mathbf{N}$ and $\mathbf{N} - \mathbf{Q_{0}}$ for the Proxima system as a function of the planet phase angle. We define the color variation as
\begin{equation}
    \Delta(X - Y) = (X - Y) - \overline{(X - Y)},
\end{equation}
where $X$ and $Y$ are respectively $\mathbf{M}$ and $\mathbf{N}$ or $\mathbf{N}$ and $\mathbf{Q_{0}}$ and $\overline{(X - Y)}$ is the mean value over the orbital period. \\
In Figure \ref{Fig_color-color}, we evaluate the color index variation for four different Proxima b orbital plane inclinations. If the planet had a face-on orbit inclination, there would be no color index variation, in contrast, approaching the edge-on inclination the color index variation increases. For Proxima Centauri System, the $\mathbf{M} - \mathbf{N}$ and $\mathbf{N} - \mathbf{Q_{0}}$ color variations are of the order of 8 ppm and 18 ppm, respectively, for an orbital plane inclination angle of $45^{\circ}$. Although these values are at the edge of the capabilities of METIS, the color variation technique could open detection possibilities for the next generation ground-based large aperture telescopes.
%

\section{Conclusions} \label{sec:conclusions}

In this work we propose and use a method to study the climate of a terrestrial exoplanet with an Earth-like atmosphere and we evaluate its detectability by photometry in the thermal IR bands. A fast and flexible 3D GCM (PlaSim) is used and modified in order to reproduce the atmosphere and the climate of the planet. 
Moreover, a robust RTC is run offline in order to determine the radiative properties of such an atmosphere. Intermediate complexity 3D GCMs have the potential to include more details compared to 1D atmospheric models, retaining enough computational speed to allow for parameter space exploration. \\
Our main conclusions are as follows:
\begin{enumerate}
\item

\emph{Proxima b climate simulation.} The recently discovered planet Proxima b is used as a case study, assuming a 1:1 gravitational resonance. 
A surface temperature distribution consistent with an open ocean day-side, and a cold and dry night-side is reported. This is in line with \cite{Boutle2017:paper} and \cite{DelGenio2019:paper}.
The permanent day-side heating is partially distributed in the night-side by the slow atmospheric circulation. Thus, in the night-side, energy is irradiated towards space in the form of IR radiation in a very efficient radiative cooling process. Consistent with other findings, e.g. \cite{Selsis2011:paper}, \cite{Showman2011:paper}, \cite{Cowan2012:paper} and \cite{Showman2013:paper}, it has been found that the warmest place of the planet surface is not located at the sub-stellar point, but it stands eastwards to that region. 
This may be due to the atmospheric superrotation associated with the planet rotation rate.
The simulations presented here, performed with an intermediate complexity model integrated with an offline RTC, are in agreement with model runs made with more sophisticated GCMs on one hand and on the other hand they allow to compute the exoplanet spectrum with a resolution of 1 nm. 
Considering the very limited computational effort, intermediate complexity models with offline RTC represent a suitable tool to perform different sensitivity studies and, at the same time, to accurately estimate the emitted and reflected planet radiations.
\item 
\emph{Detection via thermal phase curve.} 
Photometric measurements of the planet thermal emission with forthcoming instruments will be more easily exploited choosing the appropriate spectral band. This choice is not straightforward since, in each spectral band, it depends on the trade-off among the planet-to-star contrast, the amplitude of the thermal phase curve and the available photons, which affect the achievable photometric precision.
In the case of cold planets, the contrast between the IR flux of the planet and that of the star is higher for $\lambda > 10 \, \mathrm{\mu m}$. This is usually the spectral region where the amplitude of the thermal phase curve of a planet in the habitable zone is detectable with broad band photometry. However, as reported in the literature, the presence of an atmosphere mitigates the amplitude of thermal phase curve given the role of circulation in heat redistribution (see \citealt{Turbet2016:paper,Boutle2017:paper} for a more extensive description). This implies that, in similar conditions, the thermal phase curve modulation is smaller for a planet with atmosphere compared to a planet without it. Consequently, this affects the possible detection of such a planet, that is more favorable in case of the absence of an atmosphere.
Therefore, a fast model to compute the emitted IR spectrum of an exoplanet, including the effects of its atmosphere, is needed to choose the most effective spectral band in thermal phase curve observations. 
\item
\emph{Detection limits for JWST and ELT.} To understand the feasibility of a photometric detection of Proxima b, we evaluate the prospected limit of detection for the MIRIM instrument on board of the James Webb Space Telescope. Analyzing the infrared planet/star contrast for the F2100W filter of MIRIM ($\lambda = 21 \, \mathrm{\mu m}$), we find that the amplitude of the planet thermal phase curve increases by more than one order of magnitude compared to the same parameter computed for the case of the OLR. Taking into account the collecting area of JWST, MIRIM filter efficiency and an exposure time of 5 hours, we evaluate a photometric precision of $5.2 \; \mathrm{ppm}$. We also computed the photometric precision using JWST Exposure Time Calculator, obtaining, for the same exposure time, a photometric precision of $7.8 \; \mathrm{ppm}$.
Our model predicts an amplitude of Proxima b thermal phase curve that is $>15$ ppm all over the planet orbital plane inclination range $15^{\circ}> \iota > 90^{\circ}$, and therefore detectable at the level S/N $\simeq$ 1 with the aforementioned photometric precision. 
By comparison, assuming photometric shot noise, a lower limit of $\sim$ 1 hour is found for the exposure time needed for the detection in the case of ELT-METIS at S/N $\simeq$ 1. In this case the N band is considered, where the achievable photometric precision is $0.85 \; \mathrm{ppm}$, and the amplitude of Proxima b thermal phase curve is $>4$ ppm all over the planet orbital plane inclination range $15^{\circ}> \iota > 90^{\circ}$.\\
However, we have to note that in the case of JWST/MIRI this estimate is based on the assumption that total noise derives from shot noise, telescope thermal and scattered light, and from other celestial sources, \textit{e.g.}, zodiacal light, interstellar medium, and cosmic infrared background \citep{Reach1997, kelsall1998, Krick2012} as computed using the JWST ETC. This  estimate does not take into account any possible extra contribution that can potentially affect the noise floor. 
However the final performance of MIRI instrument will be fully evaluated only after the JWST commissioning, and this lead us to focus the analysis on photon-noise dominated observations. Nevertheless we briefly discuss here the expected performance of MIRI. Based on the performance of Spitzer, the noise floor for MIRI was estimated by \cite{Greene2016:paper} to be $\sim$ 50 ppm. \cite{Fauchezetal2019:paper} revised the noise floor value to a more optimistic $\sim$ 25 ppm at $1 \sigma$ confidence level, that would allow a detection for an orbital plane inclination angle $\gtrsim$ 40.
\item 
\emph{Detection via color variations.} Finally, we consider an alternative method to detect exoplanets by analyzing the color variation of an exoplanetary system in the infrared bands due to the presence of an orbiting planet. Color variation is very sensitive to the orbital inclination and to the planet temperature gradient between day-side and night-side. For a non-face-on orbit, the more the planet's hemispherical temperature difference is, the greater the color variation will be during the orbital period. For the case study of this work, the color index variation $\mathbf{M}-\mathbf{N}$ correlates with the variation $\mathbf{N}-\mathbf{Q}$. This correlation reflects the symmetry of the system under consideration. Such analyses can be very helpful to independently confirm the presence of non-transiting exoplanets analyzing the periodic color shift of a star.
\end{enumerate}
To date, because of the lack of observations in the mid infrared region with sufficient photometric precision, characterization of the atmosphere of non-transiting terrestrial exoplanets is very challenging. To foster infrared broad band photometric observations of exoplanetary systems, an improvement in photometric precision ($<10 \; \mathrm{ppm}$), at wavelengths above the $10 \; \mathrm{\mu m}$ is needed for the next generation telescopes. At the same time, the atmospheric models can pave the way for future measurement campaigns giving for instance the order of magnitude of the expected flux on telescopes, setting the observational limits and the required precision for the detection.

\section{acknowledgments}
This study was supported by grants from the Solar Physics and Space Physics group of the Department of Physics of the University of Rome Tor Vergata (https://www.fisica.uniroma2.it/$\sim$solare) and from the Atmospheric and Climate Sciences Institute (ISAC) of the Italian National Research Counsil (CNR) within the Joint Research PhD Program in Astronomy, Astrophysics and Space Science between the universities of Roma Tor Vergata, Roma Sapienza and INAF. We are also grateful to professors Nicolas Iro, Valerio Lucarini, Edilbert Kirk and Brunella Nisini for technical assistance and support. Thanks to Ilaria Giovannelli and Angela Stabile for proofreading the article.
The authors thank the anonymous reviewer for her/his valuable help in improving the manuscript.
\software{PlaSim \citep{Fraedrich2005a:paper}, libRadTran \citep{Emde2016:paper}}

\bibliography{biblio.bib}

\end{document}